%% file: main_R2.tex
\documentclass[12pt]{article}
\usepackage{amsmath}
\usepackage{graphicx}
\usepackage{enumerate}
\usepackage{url} 
\usepackage{caption}
\usepackage{subcaption}
\usepackage{natbib}
\usepackage{amsthm}
\usepackage{amsmath}
\usepackage{amsfonts}
\usepackage{amssymb}
\usepackage{color}
\usepackage{pdflscape}
\usepackage{array}
\usepackage{bbm}
\usepackage{multirow} 
\usepackage{rotating}
\usepackage{xr}
\usepackage{bm}
\usepackage{booktabs}
\usepackage{gensymb}
\usepackage{lineno}
\newcommand{\blind}{1}

\usepackage{tikz}
\usepackage{etoolbox} 
\usepackage{listofitems} 

\tikzset{>=latex} 
\colorlet{myred}{red!80!black}
\colorlet{myblue}{blue!80!black}
\colorlet{mygreen}{green!60!black}
\colorlet{mydarkred}{myred!40!black}
\colorlet{mydarkblue}{myblue!40!black}
\colorlet{mydarkgreen}{mygreen!40!black}
\tikzstyle{node}=[very thick,circle,draw=myblue,minimum size=22,inner sep=0.5,outer sep=0.6]
\tikzstyle{connect}=[->,thick,mydarkblue,shorten >=1]
\tikzset{ 
  node 1/.style={node,mydarkgreen,draw=mygreen,fill=mygreen!25},
  node 2/.style={node,mydarkblue,draw=myblue,fill=myblue!20},
  node 3/.style={node,mydarkred,draw=myred,fill=myred!20},
}
\def\nstyle{int(\lay<\Nnodlen?min(2,\lay):3)} 

\input{commands}

\begin{document}

\def\spacingset#1{\renewcommand{\baselinestretch}%
{#1}\small\normalsize} \spacingset{1}


\if1\blind
{
  \title{\bf Semi-parametric bulk and tail regression using spline-based neural networks}
  \author{
    Reetam Majumder\footnote{Corresponding Author: Reetam Majumder. Email: reetamm@uark.edu} \\
    Department of Mathematical Sciences\\ University of Arkansas\\
    and \\
    Jordan Richards \\
   School of Mathematics and Maxwell Institute for Mathematical Sciences\\ University of Edinburgh}
  \maketitle
} \fi

\if0\blind
{
  \bigskip
  \bigskip
  \bigskip
  \begin{center}
    {\LARGE\bf Semi-parametric bulk and tail regression using spline-based neural networks}
\end{center}
  \medskip
} \fi

\begin{abstract}
Semi-parametric quantile regression (SPQR) is a flexible approach to density regression that learns a spline-based representation of conditional density functions using neural networks. As it makes no parametric assumptions about the underlying density, SPQR performs well for in-sample testing and interpolation. However, it can perform poorly when modelling heavy-tailed data or when asked to extrapolate beyond the range of observations, as it fails to satisfy any of the asymptotic guarantees provided by extreme value theory (EVT). To build semi-parametric density regression models that can be used for reliable tail extrapolation, we create the blended generalised Pareto (GP) distribution, which i) provides a model for the entire range of data and, via a smooth and continuous transition, ii) benefits from exact GP upper-tails without the need for intermediate threshold selection. We combine SPQR with our blended GP to create semi-parametric quantile regression for extremes (SPQRx), which provides a flexible semi-parametric approach to density regression that is compliant with traditional EVT. We handle interpretability of SPQRx through the use of model-agnostic variable importance scores, which provide the relative importance of a covariate for separately determining the bulk and tail of the conditional density. The efficacy of SPQRx is illustrated on simulated data, and an application to U.S. wildfire burnt areas {from 1990--2020}.
\end{abstract}

\noindent%

\spacingset{1.9} 
\clearpage
\section{Introduction}\label{sec:intro}
Extreme quantile regression models provide a useful tool for risk assessment in a wide range of applied fields, such as econometrics \citep{chavez2018extreme,daouia2019extreme,daouia2023inference},  insurance \citep{daouia2023optimal}, natural hazard modelling \citep{richards2022regression,yadav2023joint}, chemometrics \citep{gardes2019integrated}, flood risk assessment \citep{johannesson2022approximate, pasche2024neural}, and downscaling \citep{wang2012estimation}. A common approach to extreme quantile regression follows via a parametric assumption on the upper-tails of the conditional distribution of a response variable $ Y \in \mathbb{R}$ given a set of realised covariates $\mathbf{x} \in \mathbb{R}^p$ of a random vector $\mathbf{X}$. The de-facto standard assumption is that excesses of $Y$ above a high threshold $u(\mathbf{x})$ follow a generalised Pareto (GP) distribution (see, e.g., \cite{coles2001introduction} for discussion) with covariate-dependent shape and scale parameters, $\xi(\mathbf{x})\in\mathbb{R}$ and $\sigma_u(\mathbf{x})>0$, respectively. That is, one assumes the parametric GP regression model:\[
\label{eq:GP_regression}
\bigg(Y-u(\mathbf{x})\bigg) \:|\;\bigg(Y > u(\mathbf{x}),\mathbf{X}=\mathbf{x}\bigg) \sim \mbox{GP}\bigg(\sigma_u(\mathbf{x}),\xi(\mathbf{x})\bigg),\]
where the scale $\sigma_u(\mathbf{x})$ is dependent on the threshold $u(\mathbf{x})$. A regression model of this form benefits from the asymptotic guarantees of the GP distribution \citep[e.g., threshold stability; see][]{coles2001introduction}, and allows for reliable estimation of extreme conditional $\tau$-quantiles (for $\tau \in (0,1)$ close to one); see \cite{davison1990models} for details.

The choice of the intermediate threshold $u(\mathbf{x})$ is an ubiquitous problem in extreme value theory (EVT) \citep[see, e.g.,][]{murphy2024automated}. In a regression setting, the threshold $u(\mathbf{x})$ can be modelled as a function of covariates. Choosing and estimating an appropriate model for $u(\mathbf{x})$ introduces uncertainty into the modelling procedure. Moreover, the additional complication that the scale $\sigma_u(\mathbf{x})$ depends on $u(\mathbf{x})$ makes interpretability of model estimates difficult \citep{richards2023insights}. To overcome the problem of threshold choice, one can replace the parametric GP model (for threshold excesses) with a model for the entire support of data that retains (exact or approximate) GP upper-tails, without requiring threshold estimation. Many such univariate models have been proposed in the literature \citep[see review by][]{scarrott2012review}. Examples include splicing or mixing of density functions \citep[with at least one constituent density being the GP density; see, e.g.,][]{frigessi2002dynamic, behrens2004bayesian, de2004data,carreau2007hybrid, reynkens2017modelling, castro2019spliced}, blending of distribution functions \citep{castro2022practical, Krakauer2024}, or parametric generalisations of the GP distribution \citep{papastathopoulos2013extended, naveau2016modeling, stein2021parametric1, stein2021parametric2, de2022extreme}. Although these all-in-one models are compliant with traditional EVT, and can thus perform reliable extrapolation into the upper-tail, they often suffer in terms of flexibility as both the bulk and tail of the distribution are parametric. Bulk and tail models with non-parametric bodies have been proposed \citep[e.g.,][]{tancredi2006accounting, macdonald2011flexible, tencaliec2020flexible}, but these are yet to be extended to the regression setting.

After estimating $u(\mathbf{x})$, a further modelling choice for GP regression is the functional forms of $\sigma_u(\mathbf{x})$ and $\xi(\mathbf{x})$. Although classical GP regression models used linear models \citep[e.g.,][]{davison1990models,  wang2012estimation}, there has been growing interest in more flexible and computationally-scalable alternatives, including additive models \citep{chavez2005generalized, youngman2019generalized}, local kernel smoothing \citep{daouia2011kernel, velthoen2019improving, gardes2019integrated}, decision trees \citep{farkas2021cyber, farkas2024generalized}, random forests \citep{gnecco2022extremalRF}, gradient boosting \citep{koh2023gradient,velthoen2023gradient}, and deep neural networks \citep{wilson2022deepgpd,richards2023insights,cisneros2024deep, pasche2024neural, mackay2024deep}. In particular, deep parametric GP regression methods \citep[i.e., those that exploit neural networks for modelling $\sigma_u(\mathbf{x})$ and $\xi(\mathbf{x})$; see overview by \cite{Richards2024}, Ch.~21 of][]{HandbookExtremes2026} are particularly appealing as they are highly flexible, scalable to high covariate dimension $p$, and can be estimated quickly using easily-accessible software packages \cite[see, e.g.,][]{pinnEV}. However, existing deep GP regression models suffer from the same two limitations of all parametric GP regression models: i) they require modelling and estimation of the intermediate threshold $u(\mathbf{x})$, and ii) they do not provide a model for the bulk (or body) and lower-tails of the conditional distribution, that is, $Y \:|\;(Y < u(\mathbf{x}),\mathbf{X}=\mathbf{x})$.

To create flexible, semi-parametric regression models that are EVT-compliant and capable of modelling the entire range of the response data, we introduce the blended GP (bGP) distribution. Similarly to the construction of the blended generalised extreme value distribution of \cite{castro2022practical}, we blend the GP distribution with an appropriately chosen distributional model for the bulk of the data. The bGP distribution can model the entire range of data, and has exact GP upper-tails without the need for threshold selection. We utilise the blended GP for regression by combining a deep GP regression model, for the upper-tails, with a deep semi-parametric regression model for the bulk of the distribution. For the latter, we employ the semi-parametric quantile regression (SPQR) model of \citet{xu-reich-2021-biometrics}, which has been used previously for inference with spatial extremal processes \citep{majumder2023deep, majumder2024modeling}. Via a spline-based representation of conditional densities, the SPQR framework provides a flexible description of the conditional bulk without any parametric assumptions. While it has been shown to work extremely well for in-sample prediction, the SPQR model assumes finite support for the response variable $Y$, and so cannot be readily used to perform extrapolation beyond the range of the observations. However, in conjunction with a blended GP regression model, we create a very flexible density regression model that is EVT-compliant. We refer to this framework as semi-parametric quantile regression for extremes (SPQRx).

The contributions of this work are threefold: i) we introduce the (unconditional) blended GP (bGP) distribution, which provides a new univariate distribution that can jointly model the bulk and upper-tail of data; ii) we develop the SPQRx framework for all-in-one, EVT-compliant density regression and extrapolation; and iii) we showcase interpretability of the SPQRx method through the use of accumulative local effects \citep[ALEs;][]{ApleyZhu2020}, which are a model-agnostic metric for assessing relative variable importance.

The rest of this paper is organised as follows. In Section~\ref{sec:SPQR}, we provide the background on SPQR. In Section~\ref{sec:XSPQR}, we introduce the blended GP distribution, and combine this with SPQR to create SPQRx, with details on estimation of variable importance provided in Section~\ref{Ss:VI_xspqr}. In Section~\ref{sec:sim_study}, we conduct simulation studies to demonstrate the efficacy of SPQRx and its ability to capture covariate effects on both the bulk and tails of the conditional density function. In Section~\ref{sec:application}, we apply SPQRx to model the entire range of burnt areas from moderate and extreme wildfire in the contiguous United States (U.S.) between 1990--2020. We conclude in Section 6 with a discussion, and outline avenues for further research.


\section{Background}\label{sec:SPQR}
\subsection{Overview}
In Section~\ref{s:spqr_model}, we provide the background details on the semi-parametric quantile regression (SPQR) framework. In Section~\ref{s:VI_spqr}, we describe accumulated local effects (ALEs) for assessing relative variable importance with SPQR.
\subsection{Semi-parametric quantile regression}\label{s:spqr_model}
Semi-parametric quantile regression (SPQR) was introduced by \citet[][]{xu-reich-2021-biometrics} as an approach to conditional density estimation that makes no parametric assumptions about the underlying distribution. Consider a response variable $ Y \in [0,1]$ and a set of realised covariates $\mathbf{x} \in \mathcal{X}$ of a random vector $\bX$, where $\mathcal{X} \subset \mathbb{R}^p$ denotes the sample space of $\bX$. Then, SPQR assumes that the conditional density of $Y \mid \bX = \bx$ is a function of $K$ third-order $M$-spline basis functions, $\{M_1(y),...,M_K(y)\}$, with $K+3$ knots spanning the unit interval $[0,1]$. $M$-spline basis functions are supported on a compact interval, positive, and integrate to one, giving them a natural interpretation as density functions. Moreover, both the $M$-spline basis function and its integral, the $I$-spline basis function, have a closed-form construction, which permits efficient inference and prediction with SPQR; see Appendix~\ref{s:spline} for details. We further note that a convex combination of $M$-spline ($I$-spline) basis functions is a valid density (distribution) function supported on a compact interval \citep{ramsay1988}. The SPQR model assumes that the conditional density can be represented as a convex combination of $M$-spline basis functions:
\begin{equation}\label{e:spqrlik}
 f_{\rm SPQR}(y|\bx) = \sum_{k=1}^Kw_k(\bx)M_k(y),
\end{equation}
with weights $w_k(\bx):\mathcal{X}\mapsto [0,1],k=1,\dots,K,$ satisfying $\sum_{k=1}^Kw_k(\bx)=1$ for all $\bx\in\mathcal{X}$.

By increasing the number $K$ of basis functions and appropriately selecting the weights, the mixture density in \eqref{e:spqrlik} can approximate any continuous density function \citep[][]{chui1980,abrahamowicz1992}. To model the weights in a flexible manner, \cite{xu-reich-2021-biometrics} use a multi-layer perceptron (MLP), a type of neural network \citep[for details, see, e.g.,][]{bishop2023deep}. The MLP has $H$ hidden layers with non-linear activation functions. Common activation functions used in previous applications of SPQR include the sigmoid and rectified linear unit \citep[ReLU,][]{NairHinton2010} functions, which we consider in this work. The additional final layer of the MLP outputs the weights $\{w_k\}_{k=1}^K$ and has a softmax activation function, which ensures that the outputs sum to one. The width of each hidden layer is determined by the number of nodes, $n_h\in \mathbb{N}$, for $h=1,\dots,H$. 

The MLP is constructed via a system of recursive equations: the output of the $h$-th layer, $\mathbf{x}^{(h)}$, is
\begin{align}\label{e:spqrnetwork}
    \bx^{(h)} &:= \boldsymbol{\phi}\left(\bW^{(h)}\bx^{(h-1)} + \bb^{(h)}\right),\quad h=1,\ldots,H,\nonumber\\
    \mathcal{W}(\bx) := \bx^{(H+1)} &= \mbox{softmax}\left(\bW^{(H+1)}\bx^{(H)} + \bb^{(H+1)}\right),
\end{align}
where $\boldsymbol{\phi}$ is the non-linear activation function, $\bx^{(0)}:=\bx$ corresponds to the input covariates, and $\mathcal{W}(\bx) = \{w_k(\bx)\}_{k=1}^K$ is a $K$-vector of weights satisfying $\sum_{k=1}^Kw_k(\bx)=1$ for all $\bx\in\mathcal{X}$. The estimable ``weights'' and ``biases'' of the MLP are contained in  $\boldsymbol{\theta} := \{\bW^{(h)},\bb^{(h)}\}_{h=1}^{H+1}$, where $\bW^{(h)} \in \mathbb{R}^{n_h \times n_{h-1}}$ and $\bb^{(h)} \in \mathbb{R}^{n_h}$ are layer-specific weight matrices and bias vectors, respectively, with input dimension $n_0:=p$. Note that the dependence of $\mathcal{W}(\cdot)$ (and its elements) on $\boldsymbol{\theta}$ has been omitted from notation for simplicity. If each layer is represented by a function, say $f^{(h)}(\cdot)$ (suppressing the dependence on $\boldsymbol{\theta}$ and $\bX$ for convenience), the MLP $f:\mathbb{R}^q \rightarrow \mathbb{R}^K$ is a composition of the layer operations, i.e., $f(\cdot) := f^{(H+1)} \circ \ldots \circ f^{(1)}(\cdot)$.

We use stochastic gradient descent and the adaptive moment estimation (Adam) optimizer \citep{kingma2014adam} to optimise $\boldsymbol{\theta}$. The optimised loss function is the negative log-likelihood associated with \eqref{e:spqrlik}. Hyper-parameters, such as the number of hidden layers $H$, the dimension $n_h$ of each hidden layer, the activation function $\boldsymbol{\phi}$, and the number of basis functions $K$, are tuned using a grid search.
Computing the conditional density function \eqref{e:spqrlik} is simple and fast given the parameters $\boldsymbol{\theta}$. Additionally, the expression for the distribution function is also available in closed form:
\begin{equation}\label{e:CDF}
 F_{\rm SPQR}(y|\bx) = \sum_{k=1}^Kw_k(\bx) I_k(y),
\end{equation}
where $\{I_k(\cdot)\}_{k=1}^K$ are $I$-spline basis functions. Note that the weights $\{w_k(\bx)\}_{k=1}^K$ do not differ between the density and distribution function in \eqref{e:spqrlik} and \eqref{e:CDF}, respectively. Finally, the conditional quantile function for level $\tau\in(0,1)$ is defined as $Q_{\rm SPQR}(\tau|\bx)$, such that 
$$F_{\rm SPQR}\{Q_{\rm SPQR}(\tau|\bx)|\bx\}=\tau.$$ While $Q_{\rm SPQR}(\tau|\bx)$ is not available in closed-form (unlike $f_{\rm SPQR}(y|\bx)$ and $F_{\rm SPQR}(y|\bx)$) it can be quickly approximated using interpolation. Finally, we note that SPQR can accommodate the special case where $\bX$ is just an intercept term, which is equivalent to unconditional density estimation; this has been recently explored by \citet{majumder2024sCDE}.
\subsection{Assessing relative variable importance for SPQR}\label{s:VI_spqr}
The SPQR framework quantifies the effect of the covariates on specific levels of the response distribution via the quantile function $Q_{\rm SPQR}(\tau|\bx)$, where $\tau \in (0,1)$ is the quantile level of interest. 
In previous work using SPQR \citep[see e.g.,][]{majumder2024modeling,majumder2023deep}, covariate effects on $\tau$-quantiles have been measured using accumulative local effects \citep[ALEs;][]{ApleyZhu2020}. We briefly outline ALEs and their link to relative variable importance quantification. We consider the effects of covariates on a generic differentiable function $g(\bx)$, where $\bx = (x_1,\ldots, x_p)$ is the vector of covariates. The sensitivity of $g(\bx)$ to covariate $x_j$ is quantified by the partial derivative 
\[
\label{eq:gdot}
\dot{g}_j(x_j) = \frac{\partial g(\bx)}{\partial x_j}.
\]
The accumulated local effect (ALE) of $x_j$ on $g(\cdot)$ is then defined as 
\begin{equation*}
    \mbox{ALE}_j(x_j; g) =\int_{z_{0,j}}^{x_j} \mathbb{E}\left[\dot{g}_j(x_j)\vert x_j = z_j\right] \mathrm{d}z_j,
\end{equation*}
where $z_{0,j}$ is an approximate lower bound for $x_j$. In practice, $\mbox{ALE}_j(x_j; g)$ is estimated by replacing the above integral with a partial sum, which can be used even when $g(\bx)$ is not differentiable. The $\mbox{ALE}_j(x_j; g)$ scores provide the relative importance of a specific value of the $j$-th covariate, $x_j$, on the function $g(\cdot)$. Following \cite{greenwell2018}, we can measure the heterogeneity of the effect of covariate $X_j$ on $g(\cdot)$ by taking the standard deviation of $\mbox{ALE}_j(X_j; g)$ with respect to the distribution of $X_j$. We thus define the variable importance (VI) score for $X_j$ on $g(\cdot)$ by 
\begin{align*}
    \mbox{VI}_j(g) &= \sqrt{\mbox{Var}_{X_j}\left[ \mbox{ALE}_j(X_j; g)\right]}.
\end{align*}
This is estimated empirically using observed samples of $X_j$. For SPQR, where $g(\cdot)$ corresponds to the conditional $\tau$-th quantile function, we typically estimate  $\mbox{VI}_j(g)$ for a sequence of $\tau$ values and for all $j=1,\dots,p$.

\section{Semi-parametric quantile regression for extremes}
\label{sec:XSPQR}
\subsection{Overview}
In Section~\ref{sec:gpd}, we present the blended generalised Pareto distribution. In Section~\ref{sec:SPQRx_details}, we describe semi-parametric quantile regression for extremes (SPQRx), which blends the SPQR representation for conditional densities with a deep GP regression model (as discussed in Section~\ref{sec:intro}). Section~\ref{Ssec:inference} details inference for SPQRx, and Section~\ref{Ss:VI_xspqr} describes estimation of relative variable importance for the conditional upper-tails.
\subsection{Blended generalised Pareto distribution}\label{sec:gpd}

\cite{castro2022practical} introduced the blended generalised extreme value distribution (bGEV), which combines the lower (exponential) tail of a Gumbel distribution with the upper (heavy) tail of a Frech\'et distribution; \cite{Krakauer2024} extended the blended GEV to allow for bounded or light-tailed margins.
The blended GEV has already been shown to be an effective model for deep extreme quantile regression; see, e.g., \cite{richards2022regression} and \cite{Richards2024}. We follow a similar approach to \cite{castro2022practical} and construct the blended GP distribution (bGP), which blends together a constituent ``bulk'' distribution, describing the body, with the GP distribution, for the upper-tails.  We present here the bGP for an arbitrary bulk distribution, in the unconditional setting, i.e., we consider a continuous random variable $Y \geq 0$, but no covariate information. We discuss the conditional setting as a natural extension towards EVT-compliant density regression in Section~\ref{sec:SPQRx_details}, where we take the bulk distribution to be the SPQR representation of conditional densities, as defined in \eqref{e:CDF}. This creates a conditional distribution function that exhibits great flexibility in the body, while retaining the exact GP upper-tails of standard parametric extreme quantile regression models.


 Let $F(y|\mathcal{W})$, $f(y|\mathcal{W})$, and $Q(y|\mathcal{W})$ denote the distribution, density, and quantile functions, respectively, of a continuous random variable, which have estimable parameters contained in $\mathcal{W}$. Here $F(y|\mathcal{W})$ and $f(y|\mathcal{W})$ are supported on the interval $[0,1]$ and referred to as the ``bulk'' distribution and density, respectively. We then let $F_{\rm GP}(y|u,\sigma_u,\xi)$, $f_{\rm GP}(y|u,\sigma_u,\xi)$, and $Q_{\rm GP}(y|u,\sigma_u,\xi)$ denote the distribution, density, and quantile functions, respectively, of a GP random variable with scale parameter $\sigma_u>0$, shape parameter $\xi\in\mathbb{R}$, and threshold $u \in \mathbb{R}$. The distribution function $F_{\rm GP}(y|u,\sigma_u,\xi)$ is
\begin{align*}
F_{\rm GP}(y|u,\sigma_u,\xi) := 
\begin{cases}
1-(1+\xi (y-u) /\sigma_u)^{-1/\xi}, \quad & \xi \neq 0, \\
1-\exp(-(y-u)/\sigma_u), \quad & \xi=0;
\end{cases}
\end{align*}
note that, unlike the aforementioned distribution for the bulk, $F_{\rm GP}(y|u,\sigma_u,\xi)$ is supported on $y \geq u$ if $\xi  \geq 0$ and, otherwise, $y \in [u, u-\sigma_u/\xi]$.

We define the blended GP distribution, denoted by bGP$(\mathcal{W},\xi),$ via its continuous distribution function $H\left(y|\mathcal{W},\xi\right)$, defined as
\begin{equation}
\label{eq:blendedH}
H\left(y|\mathcal{W},\xi\right)=\begin{cases}F(y|\mathcal{W})^{1-p(y)}F_{\rm GP}(y|\tilde{u},\tilde{\sigma}_u,\xi)^{p(y)}, \quad & y>\tilde{u}, \\
F(y|\mathcal{W}), \quad & y\leq\tilde{u}, \\
\end{cases}
\end{equation}
where $p(y)\in[0,1]$ is a weighting function. The distribution function $H(y|\mathcal{W},\xi)$ inherits support from the constituent distribution functions; the lower bound of the support is zero, whilst the upper bound is infinite if $\xi \geq 0$ and $\tilde{u}-\tilde{\sigma}_u/\xi>0$, otherwise. Following \cite{castro2022practical}, we take the weighting function to be 
\[
p(y)=p(y; a,b, c_1,c_2)=F_{\rm Beta}\left(\frac{y-a}{b-a},c_1,c_2\right),
\]
where $F_{\rm Beta}\left(\cdot,c_1,c_2\right)$ denotes the distribution function of a Beta($c_1,c_2)$ random variable with shape parameters $c_1>0$ and $c_2>0$. We blend the density functions $F(\cdot | \mathcal{W})$ and $F_{\rm GP}(\cdot | u, \sigma_u, \xi)$ within the interval $[a,b]$, where the bounds are taken to be the $p_a$ and $p_b$ quantiles of $F(\cdot)$. That is, for $p_a,p_b \in (0,1)$ with $p_a < p_b$, we take $a:=Q(p_a | \mathcal{W})$ and $b:=Q(p_b|\mathcal{W})$. To ensure continuity of $H(\cdot|\mathcal{W},\xi)$, we require $F(a|\mathcal{W})=F_{\rm GP}(a|\tilde{u},\tilde{\sigma}_u,\xi)$ and $F(b|\mathcal{W})=F_{\rm GP}(b|\tilde{u},\tilde{\sigma}_u,\xi)$. This is achieved by setting
\begin{equation}
\label{eq:sigandu}
\left(\tilde{\sigma}_u,\tilde{u}\right)=\begin{cases}
\left(\frac{\xi(a-b)}{(1-p_a)^{-\xi}-(1-p_b)^{-\xi}},a-\frac{(a-b)\{(1-p_a)^{-\xi}-1\}}{(1-p_a)^{-\xi}-(1-p_b)^{-\xi}}\right),\quad \xi \neq 0,\\
\left(\frac{(a-b)}{\log(1-p_a)-\log(1-p_b)},a-\frac{(a-b)\{-\log(1-p_a)\}}{\log(1-p_a)-\log(1-p_b)}\right),\quad \xi=0.\\
\end{cases}
\end{equation}
Note that $\tilde{u}< a$ and, for $\xi <0$, the upper-endpoint of $H(\cdot | \mathcal{W},\xi)$ satisfies $\tilde{u} - \tilde{\sigma}_u/\xi > b$; given also that $p(y) = 0$ for any $y<a$, it follows that $H(\cdot|\mathcal{W},\xi)$ is continuous. The bGP distribution has two estimable parameters, $\mathcal{W}$ and $\xi$, which respectively determine the bulk and upper-tail. The GP threshold $u$ and scale parameter $\sigma_u$ are deterministic functions of $\mathcal{W}$ and $\xi$. It is clear to see that, for large $y > \tilde{u}$, $H(\cdot|\mathcal{W},\xi)$ is exactly the GP distribution function, with a freely-varying tail index $\xi$ that is not determined by parameters $\mathcal{W}$ for the bulk distribution.

The density function of a bGP$(\mathcal{W},\xi)$ random variable has the closed-form expression
\begin{align}
\label{eq:blendedHdens}
h(y|\mathcal{W},\xi)=H(y|\mathcal{W},\xi)\times &\Bigg\{p'(y)\log F_{\rm GP}(y|\tilde{u},\tilde{\sigma}_u,\xi)+p(y)\frac{f_{\rm GP}(y|\tilde{u},\tilde{\sigma}_u,\xi)}{F_{\rm GP}(y|\tilde{u},\tilde{\sigma}_u,\xi)}\\\nonumber
&-p'(y)\log F(y|\mathcal{W}) +(1-p(y))\frac{f(y|\mathcal{W})}{F(y|\mathcal{W})} \Bigg\},
\end{align}
where $p'(y)=(b-a)^{-1}f_{\rm Beta} \left(\frac{y-a}{b-a},c_1,c_2\right)$ for $f_{\rm Beta} \left(\cdot,c_1,c_2\right)$ the density of the Beta distribution with shape parameters $c_1$ and $c_2$. As discussed by \cite{castro2022practical}, setting $c_1,c_2>3$ ensures that the second derivative of the log-density is always continuous; we hereafter assume this to hold. 

{While $h(y|\mathcal{W},\xi)$ in \eqref{eq:blendedHdens} is guaranteed to be continuous and to integrate to one, it is not guaranteed to be positive for all $y$, i.e., not a valid density function. There {exist} pathological choices of $p(\cdot)$ such that, given fixed $f(\cdot|\mathcal{W})$ and $\xi$, $h(y |\mathcal{W}, \xi)<0$ for ${y\in[a,b]}$ where $[a,b]$ is the blending interval. We note that this occurs only if $F(\cdot|\mathcal{W})$ is right-skewed and $F_{\rm GP}(\cdot| \xi)$ is relatively lighter-tailed, and $F(y|\mathcal{W})> F_{\rm GP}(y|\mathcal{W})$ for some ${y\in [a,b]}$. In practice, this is very unlikely to be an issue as $f(\cdot|\mathcal{W})$ and $\xi$ are estimated jointly, and we consider this to be only a theoretical limitation of the blended GP formulation. We further note that this is fully precluded by considering $c_2/c_1\rightarrow \infty$, which corresponds to the case where $H(y|\mathcal{W},\xi)=F(y|\mathcal{W})$ for $y\in[a,b]$ (or similarly for $c_1/c_2\rightarrow\infty$ and  $H(y|\mathcal{W},\xi)=F_{\rm GP}(y|\xi)$). Although we found no evidence of this problem occurring during inference of any models discussed hereafter, we do provide further discussion of this problem, and mitigating its occurrence during inference, in Appendix~\ref{s:valid}. }
\subsection{Tail density regression using SPQRx}
\label{sec:SPQRx_details}
We now describe semi-parametric quantile regression for extremes, SPQRx. Remaining in the unconditional setting, we first create a flexible bGP distribution by constructing the constituent bulk distribution and density functions, $F(y|\mathcal{W})$ and $f(y|\mathcal{W})$, respectively, in \eqref{eq:blendedH}, via the usual SPQR representation, i.e., we set $F(y|\mathcal{W}) = F_{\rm SPQR}(y|\mathcal{W})$ (and $f(y|\mathcal{W})=f_{\rm SPQR}(y|\mathcal{W})$), which is constructed using a basis of $I$-splines (and $M$-splines) that are supported on the interval $[0,1]$, and with basis coefficients $\mathcal{W} =\{w_k\}^K_{k=1}$ (which here are not dependent on covariates). The original SPQR method took the basis to have equally-spaced knots on $[0,1]$ \citep{xu-reich-2021-biometrics}. To reflect that SPQRx is designed for data with varying degrees of tail-heaviness (which may be covariate dependent), we instead place the knots at a sequence of quantiles with equally-spaced levels in $[0,1]$. The quantiles are estimated empirically using the training response data, and we ensure that knots are always placed at the boundaries of the unit interval.

\begin{figure}
\centering
    \includegraphics[width=0.85\linewidth]{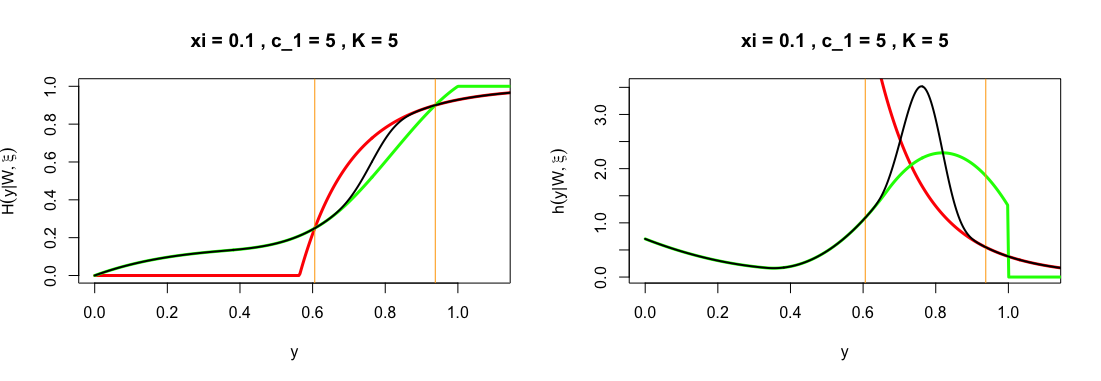}
        \includegraphics[width=.85\linewidth]{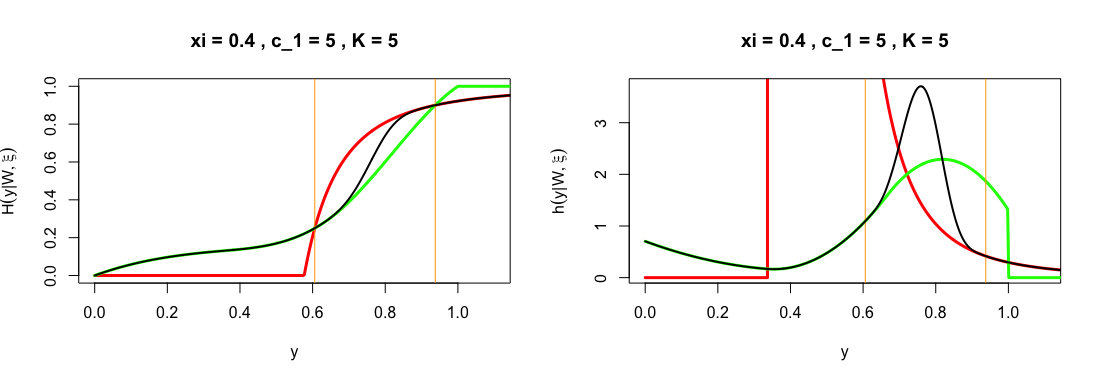}
    \includegraphics[width=.85\linewidth]{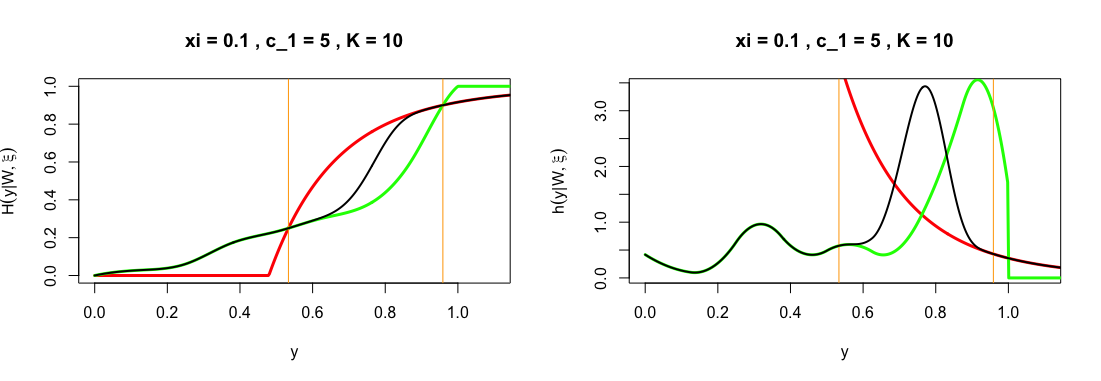}
        \includegraphics[width=.85\linewidth]{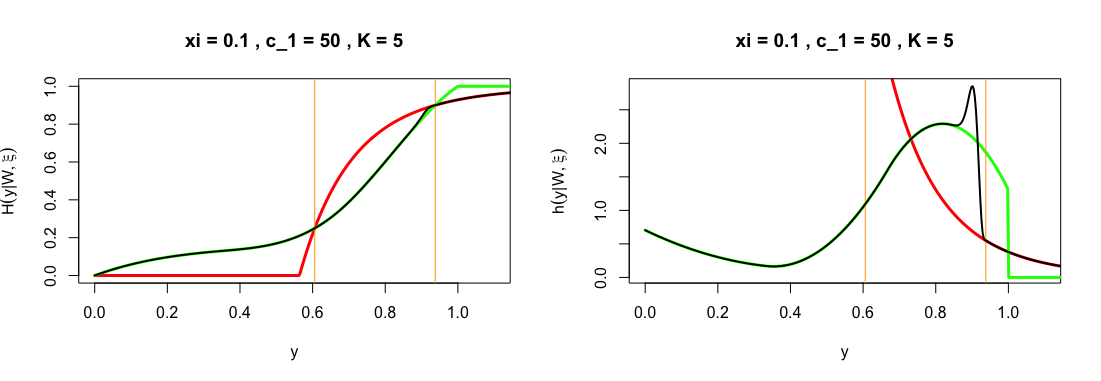}
    \caption{Illustrative examples of the blended GP distribution (left) and density (right) functions, with the constituent distribution modelled via SPQR. For a fixed random seed, we generate $K$ basis coefficients and construct the SPQR density and distribution functions (green curves). We then find $a=Q_{\rm SPQR}(p_a | \mathcal{W})$ and $b=Q_{\rm SPQR}(p_b|\mathcal{W})$, where $p_a=0.25$ and $p_b=0.9$. The values of $a$ and $b$ are denoted by the orange horizontal lines. Then, we find the required GP distribution function (red) to satisfy continuity of the bGP($\mathcal{W},\xi)$ distribution function; its resulting distribution and density functions are provided in black. The weighting function shape parameter $c_2$ is fixed to $c_2=5$ across all panels.}
    \label{fig:example}
\end{figure}

Figure~\ref{fig:example} provides illustrative examples of the distribution and density functions for this flexible SPQR-based bGP model, alongside the constituent SPQR and GP distributions and density functions. We observe great flexibility in the bulk of the distribution, and the desired exact upper GP tails. As the hyper-parameters $p_a$ and $p_b$ have a clear interpretation as the beginning and end of the blending interval, we can dichotomize the density into the ``bulk'' for $y< a$, which is purely driven by SPQR, and the ``upper-tail'' for $y \geq b$, which is driven by both SPQR and $\xi$. The hyper-parameters $c_1$ and $c_2$ determine the relative weights placed on the constituent distribution functions in \eqref{eq:blendedH} during the blending interval. Comparing the first and third rows of Figure~\ref{fig:example}, we observe that a larger value of $c_1$, relative to $c_2$, places more weight on the SPQR distribution function when blending. Given that $F_{\rm SPQR}(\cdot|\mathcal{W})$ is substantially more flexible than $F_{\rm GP}(y|\tilde{u},\tilde{\sigma}_u,\xi)$, we consider hereafter only the cases where $c_1 \geq c_2$. For simplicity, we also set $c_2=5$ throughout.

Extending the unconditional model to allow for covariate effects is straightforward. We let $Y| \mathbf{X} = \mathbf{x} \sim \mbox{bGP}(\mathcal{W}(\mathbf{x}), \xi(\mathbf{x}))$, which has conditional distribution function 
\begin{equation}
\label{eq:blend_cond_dens}
H(y|\mathbf{x})=\begin{cases}F_{\rm SPQR}(y|\mathbf{x})^{1-p(y)}F_{\rm GP}(y|\tilde{u}(\mathbf{x}),\tilde{\sigma}_u(\mathbf{x}),\xi(\mathbf{x}))^{p(y)}, \quad & y>\tilde{u}(\mathbf{x}), \\
F_{\rm SPQR}(y|\mathbf{x}), \quad & y\leq\tilde{u}(\mathbf{x}), \\
\end{cases}
\end{equation}
where $F_{\rm SPQR}(y|\mathbf{x})$ is as defined in \eqref{e:CDF}. As in \eqref{eq:sigandu}, the functions $\tilde{u}(\mathbf{x})$ and $\tilde{\sigma}_u(\mathbf{x})$ are deterministic functions of $p_a$, $p_b$, $a=Q_{\rm SPQR}(p_a |\mathbf{x})$, $b=Q_{\rm SPQR}(p_b|\mathbf{x}),$ and $\xi(\mathbf{x})$. Note that the covariates $\mathbf{x}$ enter the conditional bGP distribution through both the constituent bulk distribution function, $F_{\rm SPQR}(y|\mathbf{x}),$ and the shape parameter $\xi(\bx)$ of the GP distribution for the upper-tail. The latter is estimated alongside the weights of the conditional bulk distribution. We further note that the weighting function, $p(y)$, and its hyper-parameters, $p_a$, $p_b$, $c_1$, and $c_2$, are not dependent on covariates. 

{The SPQRx conditional quantile function, $Q_{\rm SPQRx}(\tau |\mathbf{x}):=H^{-1}(\tau|\mathbf{x})$, can be evaluated for any probability level $\tau\in[0,1]$. Although it does not have a closed form expression, its numerical evaluation is fast: for $\tau \leq p_a$, $Q_{\rm SPQRx}(\tau |\mathbf{x})=Q_{\rm SPQR}(\tau |\mathbf{x})$, which can be quickly approximated using interpolation; for $\tau \geq p_b$, $Q_{\rm SPQRx}(\tau |\mathbf{x})=Q_{\rm GP}(\tau |\tilde{u}(\mathbf{x}), \tilde{\sigma}_u(\mathbf{x}), \xi(\mathbf{x}))$ has a closed form expression; for $\tau\in[p_a,p_b]$, $Q_{\rm SPQRx}(\tau |\mathbf{x})$ must be solved numerically but the region $[p_a,p_b]$ is typically small. The continuous semi-parametric form of the conditional quantile function allows for joint estimation of quantiles at a sequence of levels $\tau$. Thus, as long as $Q_{\rm SPQRx}$ is monotonically increasing, our method {does not suffer} from the crossing problems of standard quantile regression approaches \citep[see, e.g.,][]{saleh2021solution}. }

As outlined in \eqref{e:spqrnetwork}, the basis weights $\mathcal{W}(\mathbf{x})=\{w_k(\mathbf{x})\}_{k=1}^K$ of the conditional density function in the SPQR framework, $f_{\rm SPQR}(y|\mathbf{x})$, are modelled as the final layer of a multi-layer perceptron. A softmax final activation layer is applied to ensure that the $K$-vector of weights must sum to one. For SPQRx, we must also model the shape parameter function, $\xi(\mathbf{x})$. A natural and parsimonious choice is to combine the modelling of $\xi(\mathbf{x})$ with that of $\mathcal{W}(\mathbf{x})$. To this end, we jointly model $(\xi( \bx),\mathcal{W}(\bx))$ as the final layer of a multi-layer perceptron (see Figure~\ref{fig:schematic}) with an appropriately chosen final activation layer. We replace the final layer of \eqref{e:spqrnetwork} by
$$
 \left(\xi( \bx),\mathcal{W}(\bx)\right) = \bx^{(H+1)} = \mbox{softmax}^*\left(\bW^{(H+1)}\bx^{(H)} + \bb^{(H+1)}\right),
$$
where the $\mbox{softmax}^*(\cdot)$ activation function applies a $\xi$-specific activation function to the first element of its input vector, and the usual $\mbox{softmax}(\cdot)$ activation to the last $K$ elements. The $\xi$-specific activation function should be chosen to ensure an appropriate range for $\xi(\mathbf{x})$, with examples including: exponential ($\xi(\mathbf{x})>0$), sigmoid ($\xi(\mathbf{x}) \in (0,1)$), or tanh ($\xi(\mathbf{x})\in (-1,1)$); see discussion by \cite{Richards2024}.
\begin{figure}[!t]
  \centering
\begin{tikzpicture}[x=2.8cm,y=1.4cm]
  \readlist\Nnod{3,4,4,3} 
  \readlist\Nstr{p,n_{\prev},} 
  \readlist\Cstr{x,x^{(\prev)},\nu(\mathbf{x})} 
  \def\yshift{0.55} 
  \foreachitem \N \in \Nnod{
    \def\lay{\Ncnt} 
    \pgfmathsetmacro\prev{int(\Ncnt-1)} 
    \foreach \i [evaluate={\c=int(\i==\N); \y=\N/2-\i-\c*\yshift;
                 \x=\lay; \n=\nstyle;
                 \index=(\i<\N?int(\i):"\Nstr[\n]");}] in {1,...,\N}{ 
       \ifnum \lay<\Nnodlen
         \node[node \n] (N\lay-\i) at (\x,\y) {$\strut\Cstr[\n]_{\index}$};
      \fi
       \ifnum \lay=\Nnodlen
       \ifnum \i=1
         \node[node \n] (N\lay-\i) at (\x,\y) {$\xi(\mathbf{x})$};
         \fi
           \ifnum \i=2
         \node[node \n] (N\lay-\i) at (\x,\y) {$w_1(\mathbf{x})$};
         \fi
           \ifnum \i=3
         \node[node \n] (N\lay-\i) at (\x,\y) {$w_K(\mathbf{x})$};
         \fi
      \fi
      
      \ifnumcomp{\lay}{>}{1}{ 
        \foreach \j in {1,...,\Nnod[\prev]}{ 
          \draw[white,line width=1.2,shorten >=1] (N\prev-\j) -- (N\lay-\i);
          \draw[connect] (N\prev-\j) -- (N\lay-\i);
        }
        \ifnum \lay=\Nnodlen
          \draw[connect] (N\lay-\i) --++ (0.5,0); 
        \fi
      }{
        \draw[connect] (0.5,\y) -- (N\lay-\i); 
      }
      
    }
      \ifnum \lay<\Nnodlen
        \path (N\lay-\N) --++ (0,1+\yshift) node[midway,scale=1.6] {$\vdots$}; 
        \fi
         \ifnum \lay=\Nnodlen
        \path (N\lay-\N) --++ (0,.8+\yshift) node[midway,scale=1.6] {$\vdots$}; 
        \fi
}
 
  \node[above=3,align=center,mydarkgreen] at (N1-1.90) {Input\\[-0.2em]layer};
  \node[above=2,align=center,mydarkblue] at (N2-1.90) {Hidden\\[-0.2em]layer 1};
  \node[above=2,align=center,mydarkblue] at (N3-1.90) {Hidden\\[-0.2em]layer 2};
  \node[above=3,align=center,mydarkred] at (N4-1.90) {Output\\[-0.2em]layer};
 
\end{tikzpicture}
\caption{Schematic of the underlying multi-layer percepton (MLP) of an SPQRx model with $H=2$ hidden layers. The input covariates are $\mathbf{x} =(x_1,\dots,x_p)$, and the output of the final $\mbox{softmax}^*(\cdot)$ layer is $(\xi(\mathbf{x}),w_1(\mathbf{x}),\dots,w_K(\mathbf{x})),$ where $w_1(\mathbf{x}),\dots,w_K(\mathbf{x})$ are the $K$ basis functions comprising the constituent SPQR density/distribution.}
\label{fig:schematic}
\end{figure}

 {Throughout}, we refer to this methodology of density regression as SPQR for extremes, or SPQRx. The nomenclature emphasises the flexible distributional assumptions for our density regression model, but we note that alternative models for the body of the data could also be used in combination with the blended GP.

\subsection{Inference}
\label{Ssec:inference}
 Inference for SPQRx follows similarly to SPQR, as described in Section~\ref{s:spqr_model}. We use Adam, with default hyper-parameters, to train the MLP that defines the parameter vector $(\xi(\mathbf{x}),w_1(\mathbf{x}),\dots,w_K(\mathbf{x}))$, where the loss function is the negative log-likelihood associated with the conditional bGP density in \eqref{eq:blend_cond_dens} {and a penalty term to encourage positivity of the conditional density function in the blending intervals (see discussion in Section~\ref{sec:gpd} and Appendix~\ref{s:valid}).} We note that the hyper-parameters of the blending function $p(y)$ in \eqref{eq:blendedH} are fixed during training of the model. Prior to model fitting, observations $\{(y_i,\mathbf{x}_i)\}_{i=1}^n$ are split into training and test sets, $\{(y_i,\mathbf{x}_i)\}_{\rm train}$ and $\{(y_i,\mathbf{x}_i)\}_{\rm test}$, respectively. We assume our observations $y_i$ satisfy $y_i>0,i=1,\dots,n$. However, as the SPQR density is supported on the unit interval, and training response observations $\{y_i\}_{\rm train}$ may not fall in $[0,1]$, we must first standardise $\{y_i\}_{\rm train}$ to the unit interval via a min-max transformation; this is based on the training data alone. We use the same values to standardise $\{y_i\}_{\rm test}$, however, there are no guarantees that all observations of the resulting standardised test response data will satisfy $y_i<1$. Although this is not an issue for SPQRx, any such test observation cannot be handled by SPQR without re-training of the model, and all associated probability and density estimates will degenerate to one and zero, respectively. Examples of this phenomena are illustrated in the application, Section~\ref{sec:application}.

The SPQRx model is trained using the \texttt{R} interface to \texttt{keras} \citep{kerasforR}. Before training, we further hold out $20\%$ of the training data for validation during optimisation of the MLP. To mitigate overfitting, our training scheme uses checkpoints and early-stopping \citep{prechelt2002early}. At each iteration of the optimisation algorithm, we evaluate the negative log-likelihood on the validation data, and then determine the ``best fitting'' model (over all iterations) as that which minimises the validation loss function. Early-stopping halts the optimisation scheme early if the validation loss has not decreased in the last $\Delta\in\mathbb{N}$ iterations (we set $\Delta = 25$ throughout). We also add $L_1$ penalties to the final activation layer for the shape parameter, $\xi(\mathbf{x})$, which provides further model regularisation; see, e.g., \cite{goodfellow2016deep}. 

Courtesy of the modular nature of the SPQRx density, we can use pre-training (see \cite{goodfellow2016deep}, Ch.8) to increase the  computational efficiency of the inference procedure and mitigate numerical instability during training. When fitting the SPQRx model, we first fit the corresponding SPQR model, i.e., with $p_a=p_b\rightarrow 1$. We then use the parameters of the estimated neural network as initial estimates when training SPQRx. We find that this procedure works well in practice.

While we presented the SPQRx method in Section~\ref{sec:XSPQR} for a real-valued shape parameter, $\xi(\mathbf{x})\in\mathbb{R}$, there may be computational and practical benefits to constraining $\xi(\mathbf{x})$ to be non-negative. When the shape parameter $\xi(\mathbf{x})$ is negative, the resulting GP distribution has a finite upper endpoint, which can make validation and testing of GP regression models computationally troublesome. If test or validation response data are observed to exceed their predicted upper-endpoint, then the negative log-likelihood function will evaluate to a non-finite value. As a consequence, the loss surface is highly non-regular, and iterative stochastic gradient descent methods (e.g., Adam) may then struggle to find global maxima. For a discussion of this issue, particularly with respect to deep GP regression models, see \cite{richards2022regression, Richards2024}.  {If negative values of $\xi(\mathbf{x})$ are desirable, the training scheme of \cite{mackay2024deep} can be employed. They advocated initialising the model with $\xi(\mathbf{x})$ strictly positive for all $\mathbf{x}$. The loss function is thus guaranteed to be finite at the onset of training. Then, if the optimisation algorithm produces non-finite loss values during training, training restarts from the last iteration (with a finite loss) and the learning rate is reduced. This procedure is repeated for a sequence of decreasing learning rate values. }

 \subsection{Relative variable importance for SPQRx}\label{Ss:VI_xspqr}
 As with many applications of density regression models, we seek to assess the relative importance of covariates on determining the conditional distribution. In the SPQRx framework, the parameters of the conditional density comprise those that influence the bulk (i.e., $\mathcal{W}(\mathbf{x})$) and those that are specific to the tails (i.e., $\xi(\mathbf{x})$). Thus, SPQRx permits the covariates to separately affect the body and the tail of the conditional distribution. As a result of this, we may expect the relative importance of covariates to change between the bulk and tails. 

 Following the discussion in Section~\ref{s:VI_spqr}, we assess relative importance in the bulk by estimating the VI scores for a sequence of $\tau$ values; recall that, for $\tau < p_a$, the conditional distribution function for the SPQRx model is exactly that of the usual SPQR framework. When considering relative importance in the tails, we have two options: i) considering VI scores for $\tau$ close to one or ii) directly interpreting relative importance for the shape parameter, $\xi(\mathbf{x})$. The latter can be achieved similarly to the conditional quantile function (for fixed $\tau$), by setting $g(\mathbf{x})$ in \eqref{eq:gdot} to be $\xi(\mathbf{x})$.

\section{Simulation Study}
\label{sec:sim_study}
\subsection{Overview}
Here we provide a simulation study to showcase the benefits of using SPQRx over SPQR {and competing deep quantile regression methods,} when extrapolating into the tails. Covariates $\mathbf{X}_i, i=1,\dots,3$, are drawn independently from a Unif(0,1) distribution. We then draw our response $Y\mid(\mathbf{X}=\mathbf{x})$ from a log-normal distribution with density function $f(y | \mathbf{x})=(\sqrt{2\pi}\sigma(\mathbf{x}) y)^{-1}\exp\{-(\log y - \mu(\mathbf{x}))^2/(2\sigma^2(\mathbf{x}))\}$ for $y>0, \sigma(\mathbf{x}) >0,$ and $\mu(\mathbf{x}) \in \mathbb{R}$. The parameters $\mu(\mathbf{x})$ and $\sigma(\mathbf{x})$ are dependent on $\mathbf{x},$ and we set $\mu(\mathbf{x})=  5(1-1/(1+\exp(-(1-5x_1x_2))))$ and $\sigma(\mathbf{x})=1/(1+\exp(-(1-5x_1x_2)))$; hence only $X_1$ and $X_2$ act on $Y$.
 
To evaluate the efficacy of conditional density estimation, we compute the integrated conditional 1-Wasserstein distance (IWD), defined by
\[\label{IWD}
\mbox{IWD}=\int_\mathcal{X}\int^1_0 \left| Q(y | \mathbf{x})-\hat{Q}(y|\mathbf{x})\right|\mathrm{d}\mathbf{x},
\]
where $Q(y|\mathbf{x})$ denotes the true conditional quantile function (and $\hat{Q}(y|\mathbf{x})$ denotes its estimate), and where $\mathcal{X}$ denotes the sample space for the covariate vector $\mathbf{X}$ (in the above case, $\mathcal{X}=[0,1]^3$). We estimate the out-of-sample IWD by generating a test set of 5000 covariate vectors and performing Monte-Carlo integration. We also consider a tail-calibrated version of the IWD, denoted by tIWD, which is constructed by replacing the limits of the inner integral of \eqref{IWD} with $[0.999,1]$.

We perform 100 experiments. For each, we generate $n \in \{1000,10000\}$ samples from the aforementioned log-normal regression model, and fit the original SPQR model, the SPQRx model, and {a deep GP regression model \citep{wilson2022deepgpd, richards2023insights}, using the parameterisation proposed by \cite{pasche2024neural}. The deep GP model involves a two-step estimation procedure: first, a deep quantile regression model is used to estimate a non-stationary exceedance threshold, corresponding to the conditional $p_a$-quantile (for $p_a$ close to one). We then fit a non-stationary deep GP model to excesses above this threshold. The intermediate exceedance quantile and the deep GP parameters are represented as MLPs with the same architecture (i.e., number of layers and width per layer). Note that the deep GP regression model can only describe the conditional quantile function, $Q(y|\mathbf{x}),$ for $y \geq Q(p_a|\mathbf{x})$. Thus, we cannot evaluate the IWD for this model.} 

For all three models, we consider a coarse grid of hyper-parameters which is motivated by unreported initial experiments. For the $M$-spline basis of the SPQR density, we consider $K\in\{15,25\}$ basis functions. {For fairness of comparison, 
we consider the same underlying architectures for all aforementioned MLPs (SPQR, SPQRx, and the two that comprise the deep GP model): a $2$-layered neural network with $n_h\in\{16,32\}$ nodes per layer, and two activation functions $\boldsymbol{\phi}(\cdot)$: sigmoid and ReLU.} For the hyper-parameters specific to SPQRx, we consider $p_a \in \{0.75,0.9\}$, $p_b\in\{0.99,0.999\},$ and $c_1 \in \{5,25\}$. {For the deep GP regression model, we consider $p_a \in \{0.75,0.9, 0.925\}$. For both SPQRx and the deep GP model, the $\xi$-specific activation function is taken to be the hyperbolic tangent function, scaled and shifted to ensure that $\xi(\mathbf{x}) \in (-0.5,0.7)$ \citep[see][]{pasche2024neural}. In both cases, we use the iterative training scheme described in Section~\ref{Ssec:inference} to accommodate the negative shape values. }

\subsection{Results}

\begin{sidewaystable}
\centering
 \caption{Results of the simulation study. The median of the IWD and tIWD estimates (alongside 50\% confidence intervals; in brackets) are reported for the {original SPQR, new SPQRx model (in bold), and the deep GP regression model (underlined). Note that the IWD is undefined for the deep GP model and the hyperparameter $K$ is not used in its construction}.  The hyper-parameters of the SPQRx, $(p_a,p_b,c_1)$, are optimised for each row by minimising the tIWD, and are reported in the final column. }
 \label{table}
\begin{tabular}{rrrccr}
  \hline
$n$ & $K$ & $n_h$ & IWD & tIWD & Optimal $(p_a,p_b,c_1)$\\ 
  \hline
\multirow{4}{*}{1000} & 15 & 16 & 2.93 (2.74, 3.29)/\textbf{2.84 (2.63, 3.31)} & 9.35 (8.06, 11.3)/\textbf{9.89 (8.64, 11.4)} & $(0.9,0.999,5)$\\ 
& 15 & 32 & 3.97 (3.51, 4.24)/\textbf{3.81 (3.27, 4.16)} & 9.24 (8.34, 10.6)/\textbf{9.23 (8.23, 10.6)} & $(0.9,0.999,5)$\\ 
   & 25 & 16 & 5.75 (5.63, 5.94)/\textbf{2.96 (2.83, 3.12)} & \shortstack{12.0 (11.2, 13.0)/\textbf{9.56 (8.42, 10.9)} \\ \underline{16.1 (13.6, 20.5)}} & $(0.9,0.999,5)$\\ 
   & 25 & 32 & 3.50 (2.39, 4.09)/\textbf{3.28 (2.24, 3.90)} & \shortstack{9.02 (7.99, 10.3)/\textbf{9.11 (8.09, 10.3)}\\ \underline{17.6 (14.5, 22.4)}} & $(0.9,0.999,5)$\\ 
   \hline
  \multirow{4}{*}{10000}  & 15 & 16 & 1.85 (1.80, 1.91)/\textbf{1.79 (1.75, 1.85)} & 10.4 (9.22, 11.4)/\textbf{8.30 (7.46, 9.34)} & $(0.75,0.99,25)$\\ 
   & 15 & 32 & 1.87 (1.81, 1.93)/\textbf{1.83 (1.77, 1.89)} & 10.9 (10.1, 11.8)/\textbf{9.62 (8.79, 10.3)}  & $(0.75,0.99,25)$\\ 
   & 25 & 16 & 1.01 (0.95, 1.07)/\textbf{0.91 (0.86, 0.96)}& \shortstack{8.73 (7.38, 10.6)/\textbf{6.16 (4.92, 7.41)}\\\underline{ 9.25 (7.98, 11.7)}} & $(0.9,0.99,25)$\\  
   & 25 & 32 & 1.03 (0.98, 1.11)/\textbf{0.91 (0.86, 0.95)} & \shortstack{10.1 (9.26, 11.2)/\textbf{6.55 (5.69, 7.52)} \\ \underline{9.45 (8.36, 11.3)}} & $(0.9,0.99,25)$\\ 
   \hline
\end{tabular}
\end{sidewaystable}

Table~\ref{table} provides the results of the simulation study. For brevity, we present here only the results when $\boldsymbol{\phi}(\cdot)$ in \eqref{e:spqrnetwork} is the sigmoid activation, which  provides consistently better results compared to the ReLU activation function. {We also present only the optimal results for the deep GP regression model.} We observe that SPQRx generally provides lower IWD and tIWD estimate than the competing SPQR method, which illustrates the efficacy of SPQRx for modelling heavy-tailed conditional densities. {The SPQRx model consistently outperforms the deep GP regression model, and further benefits from a full description of the conditional distribution not offered by peaks-over-threshold models.} As one might expect, the estimation accuracy tends to increase with the number of knots and the number of nodes in the underlying MLP. For the optimal hyper-parameters of SPQRx, we find a lower $c_1$ is preferred in a low sample setting, whilst a higher $c_1$ is preferred when more data are available for inference. As discussed alongside Figure~\ref{fig:example}, a higher value of $c_1$ puts more weight on the more flexible SPQR distribution during blending. This may be preferable when you have more data available to infer this component of the model. For the blending interval, we find consistent agreement, when $n=1000$, that the preferred blending quantile levels are $p_a=0.9$ and $p_b=0.99$. This provides a relatively wide interval for blending of the two constituent distribution functions, with the exact GP upper-tails beginning at $b$. Although Table~\ref{table} provides some intuition as to appropriate choices of hyper-parameters for SPQRx, it is by no means conclusive. We advocate that, in practice, a grid-search over hyper-parameters is performed and the quality of SPQRx fits are compared on hold-out data.
\begin{figure}[ht!]
\centering
    \includegraphics[width=0.32\linewidth]{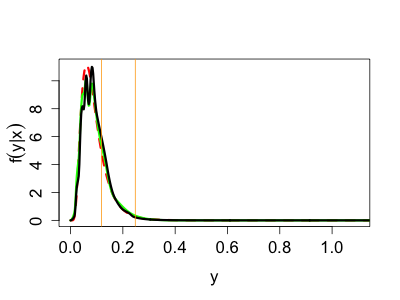}
        \includegraphics[width=0.32\linewidth]{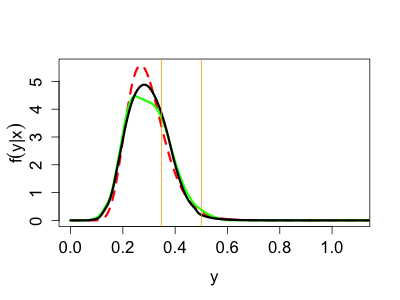}
            \includegraphics[width=0.32\linewidth]{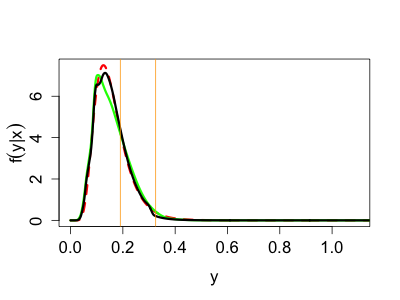}
        \includegraphics[width=.32\linewidth]{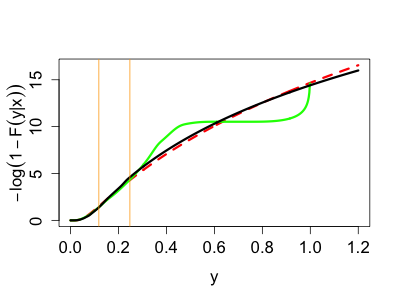}
    \includegraphics[width=0.32\linewidth]{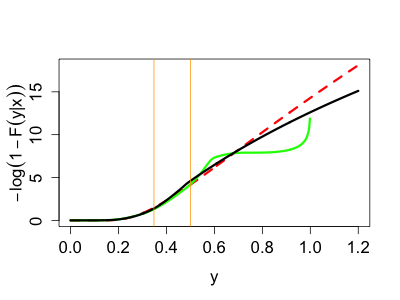}
        \includegraphics[width=.32\linewidth]{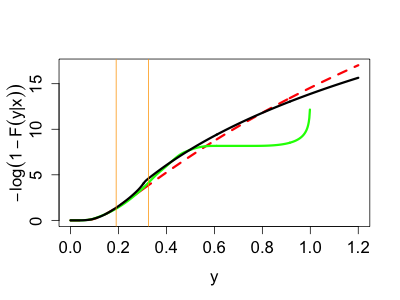}
 
    \caption{Simulation study: estimates of  density (top) and log-survival (bottom) functions for three test covariate vectors. The red dashed lines give the true functions. The black and green curves are the corresponding estimates from SPQR and SPQRx, respectively. The values of $a$ and $b$ (from the SPQRx fit) are denoted by the orange horizontal lines.}
    \label{fig:simstudy1}
\end{figure}
 \begin{figure}[ht!]
\centering
         \includegraphics[width=.65\linewidth]{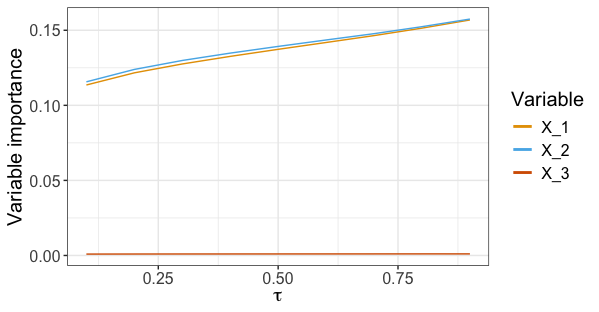}
    \caption{Simulation study: estimates of the  variable importance scores $\mbox{VI}_j\left(\hat{Q}(\tau|\mathbf{x})\right)$ for the covariates $X_j,j=1,2,3,$ as a function of the quantile level $\tau$.}
    \label{fig:simstudy2}
\end{figure}

Figure~\ref{fig:simstudy1} compares out-of-sample density and distribution estimation using SPQR and SPQRx. In this case, we take model fits to $n=10000$ samples, using the optimal hyper-parameters from the bottom row of Table~\ref{table}. As Figure~\ref{fig:simstudy1} illustrates, SPQRx is much better able to capture the upper-tails of the true log-normal distribution function. Moreover, SPQR is unable to provide any useful estimates of the conditional distribution function for $y>1$. Figure~\ref{fig:simstudy2} also provides estimates of the variable importance scores from the aforementioned fitted SPQRx model. Specifically, we estimate $\mbox{VI}_j\left(\hat{Q}(\tau|\mathbf{x})\right)$ where the function of interest $g(\cdot)$, defined in \ref{eq:sigandu}, is taken to be the estimated conditional $\tau$-quantile function.  The estimates are able to capture the true underlying covariate importance, specifically that $X_3$ does not have any effect on the conditional distribution at any quantile level $\tau$, whilst $X_1$ and $X_2$ have an equal effect for all values of $\tau$. The $\xi$-specific variable importance scores (see Section~\ref{Ss:VI_xspqr}) are in agreement, with estimates $0.0156, 0.0234, 0.0009$ for $X_1$, $X_2$, and $X_3$, respectively.

 {Two additional simulation studies are described in Appendix~\ref{s:sim_studies}. Here we consider the higher-dimensional setting, where $p=20$, and where the heaviness of the conditional tails is allowed to vary with the covariates (i.e., $\xi(\mathbf{x}) \neq 0$ for all $\mathbf{x}$). We find that SPQRx outperforms SPQR in both studies. When data are heavy-tailed, SPQRx also outperforms the deep GP model, but performs slightly worse when the data have bounded upper-tails, i.e., $\xi(\mathbf{x})<0$ for all $\mathbf{x}$. In this case, estimation of the SPQRx model can be computationally troublesome (see Section~\ref{Ssec:inference}). In our application to U.S. wildfire burnt areas, the data are demonstrably heavy-tailed.}
\section{Application to U.S. wildfire burnt areas}
\label{sec:application}
We demonstrate the flexibility of SPQRx by modelling U.S. wildfire burnt areas, that were originally compiled and analysed by \citet{lawlershaby2024}. The data consist of burnt areas (in 1000s of acres) and counts for over 10,000 wildfires across the contiguous U.S., from 1990 to 2020, with only fires exceeding 1000 acres of burnt area included in the dataset. The dataset includes several covariates, including spatio-temporal coordinates, meteorological variables, fire weather indices, and housing density. \citet{lawlershaby2024} studied these data using a joint model for counts and area with spatial random effects, and employed the extended GP distribution for modelling burnt areas. In our work, we focus on modelling the distribution of the continuous burnt areas, conditional on the following 6 covariates: year (\texttt{fire\_yr}), total precipitation during the month of the fire (\texttt{pr\_curr}; mm) and during the entire previous year (\texttt{pr\_prev}; mm), minimum relative humidity (\texttt{rmin}; $\%$-age), maximum temperature (\texttt{tmax}; K), and windspeed (\texttt{wspd}; m/s). The final three variables (\texttt{rmin, tmax, wspd}) are aggregated into their monthly means.

Following \cite{lawlershaby2024}, we split the data into training and testing sets of 6416 and 3344 fires, respectively. The split is not random, as the test set consists of the first and last five years of the data: 1184 fires from 1990 to 1994, inclusive, and 2160 fires from 2016 to 2020, inclusive. Additionally, we uniformly-at-random select $20\%$ of the training data for validation during training of the SPQRx model; see Section~\ref{Ssec:inference}. As illustrated by \cite{richards2022regression} and \citet{lawlershaby2024}, U.S. wildfire burnt areas are extremely heavy-tailed ($\xi > 0.5$), which can cause numerical difficulties during training of deep extreme quantile regression models \citep{richards2022regression}. While \citet{lawlershaby2024} employ a log-transform of the data to mitigate this issue, we instead follow \cite{richards2022regression} and model the square-root of the burnt areas as our response variable, $Y$. The square-root transformation better preserves the tail behavior of the response compared to the log-transform, and can be interpreted as the diameter of an affected region. The wildfires are assumed to be independent of each other; while a naive assumption, it is nevertheless aligned with our specific goal of testing the appropriateness of the SPQRx framework for density regression.

We fit and compare both the original SPQR and the EVT-compliant SPQRx models to the data. When deciding on the optimal hyper-parameters for the SPQR/SPQRx model, we perform a coarse grid search over hyper-parameters to choose their optimal values. We use the same MLP architecture for both frameworks. The candidate models are compared using the negative log-likelihood (evaluated on the test data) and by visual examination of a model diagnostic, the pooled QQ plot, which is to be discussed. The optimal neural network comprising our SPQR/SPQRx model has $H=2$ hidden layers, each with $n_h=12$ nodes, and with $K=25$ basis functions describing the body of the distribution.  An exploratory data analysis, whereby we fit stationary GP models to the response data, reveals that the response burnt area data are in fact heavy-tailed, with the vast majority of fits returning estimates of $\xi$ exceeding zero. {For simplicity, we thus constrain $\xi(\mathbf{x})>0$ for SPQRx by using the logistic activation function. This avoids the need for the burdensome training scheme described in Section~\ref{Ssec:inference}.} For the blending interval, we set $p_a$ and $p_b$ to be $0.9$ and $0.999$, respectively. Finally, $c_1$ and $c_2$ are set to be $25$ and $5$, respectively, with these two values being carried over from our numerical studies. {Most of these settings are chosen based on an initial unreported coarse grid-search over hyper-parameters. We anticipate that relaxing the assumption of $\xi(\bx)>0$, or choosing $(c_1,c_2)$ based on the grid search, could change the `optimal' architecture for our analysis.}

 The covariates for both the training and test sets are normalised by subtracting and dividing by the marginal mean and standard deviation vectors, respectively, of the training covariate data. Similarly, the response data are scaled to $[0,1]$ using the min-max transformation, as described in Section~\ref{Ssec:inference}. As a consequence, there are two response values in the scaled test data that exceed one, corresponding to the two largest wildfires across the entire dataset, which occurred in 2017 and 2020. As explained in Section~\ref{Ssec:inference}, regular SPQR can not provide any usable estimates for these particularly extreme test values, but this is not an issue for SPQRx. 
 {Estimation uncertainty is quantified via 200 bootstrap samples of the training period data. Note that the validation data change with every bootstrap sample.}

\begin{figure}[t]
    \centering
    \includegraphics[width=0.8\linewidth]{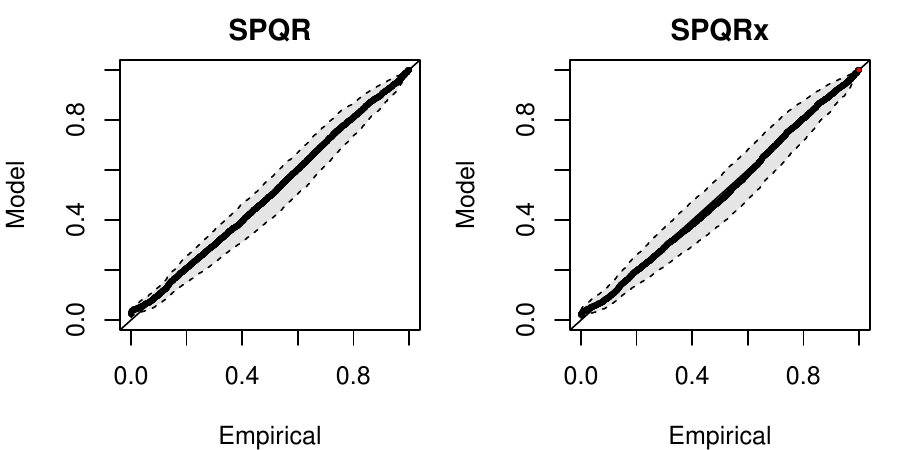}
    \includegraphics[width=0.8\linewidth]{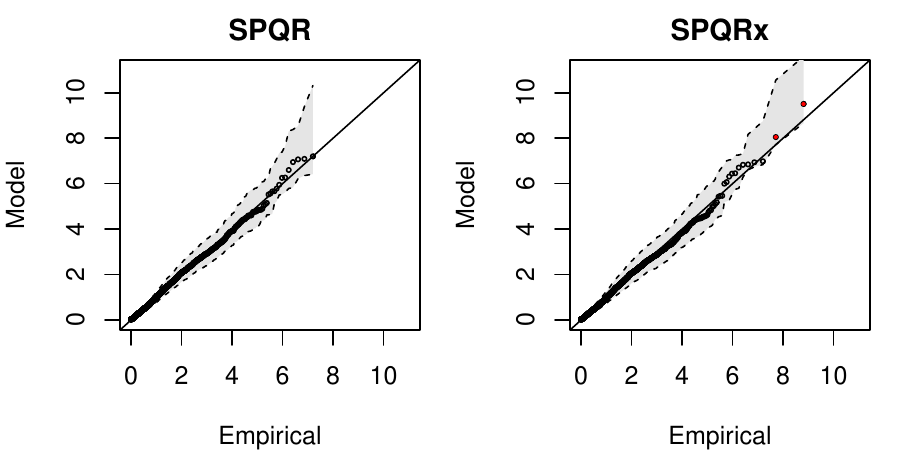}
    \caption{Goodness-of-fit diagnostics for the SPQR and SPQRx models. Pooled PP (top) and QQ (bottom) plots for the fitted SPQR (left) and SPQRx (right) models, evaluated on the test data. The red points in the right column denote the two most extreme test values, which cannot be modelled using SPQR and thus do not appear on the left column. {The grey area bounded by the dashed lines represent the $95\%$ bootstrap confidence intervals}.}
    \label{fig:ff_qqplot_exp}
\end{figure}

Figure~\ref{fig:ff_qqplot_exp} provides visual goodness-of-fit diagnostics for the fitted SPQR and SPQRx models. We create out-of-sample pooled PP and QQ plots, whereby the fitted conditional models are used to transform the test data onto standard (unconditional) margins. The top row of PP plots are on uniform margins which showcase goodness-of-fit for the entire distribution, while the bottom row QQ plots are on exponential margins, to highlight the fits in the upper-tails \citep[see, e.g.,][]{heffernan2001extreme}; deviation of points from the diagonal suggest poor fits. {Both SPQR and SPQRx provide fantastic fits for the entire range of the data. However, SPQRx benefits from an asymptotically-justified mechanism for extrapolating into the upper-tails. This is illustrated by the red points in Figure~\ref{fig:ff_qqplot_exp}, which correspond with the two most extreme burnt areas in the test data. While SPQRx is capable of predicting well these events, SPQR is unable to provide any usable probability statements, as they are outside of the range of the training data (see Section~\ref{Ssec:inference}). }



\begin{figure}[t!]
    \centering
    \includegraphics[width=\linewidth]{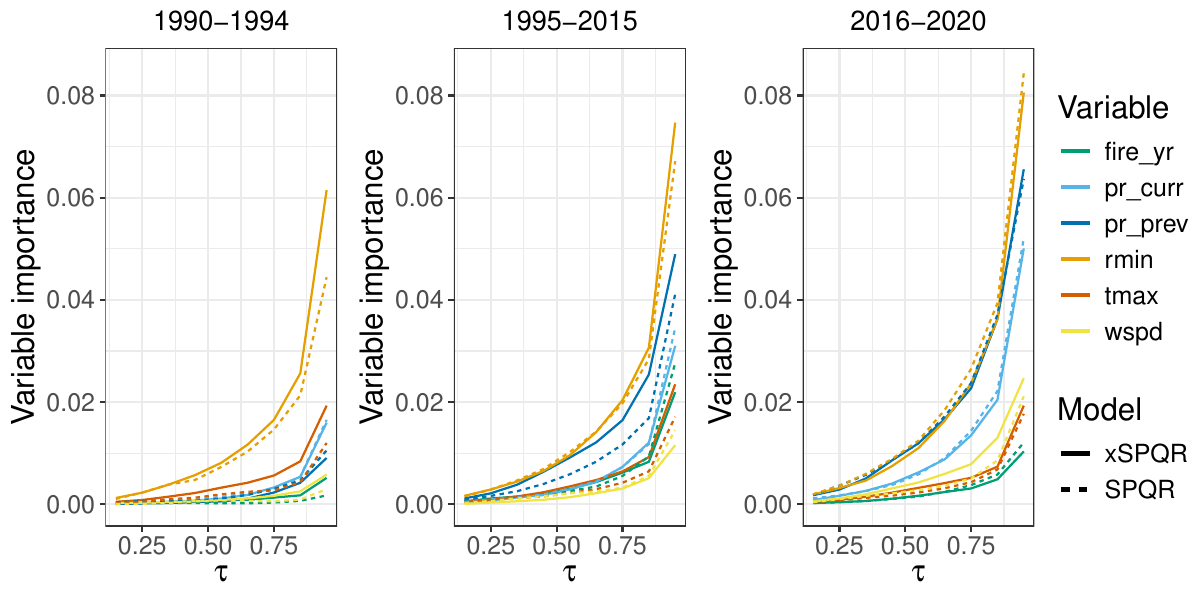}
    \caption{Estimates of the  variable importance scores $\mbox{VI}_j\left(\hat{Q}(\tau|\mathbf{x})\right)$ as a function of the quantile level $\tau$. Different colours correspond to different covariates, while the dashed and solid lines correspond to estimates from SPQR and SPQRx, respectively. The left and right panels correspond to the two test periods: 1990--1994 and 2016--2020, respectively. The central panel corresponds to the training period, 1995--2015. }
    \label{fig:ff_VI}
\end{figure}

Figure~\ref{fig:ff_VI} provides estimates of the variable importance scores for both the SPQR and SPQRx model fits. We consider three time periods: the two test periods, 1990--1994 and 2016--200, and the training period, 1995-2015. Across both models and all time periods, minimum relative humidity \texttt{rmin} is identified as the most important variable for predicting the distribution of U.S. wildfire burnt areas across all quantile levels, $\tau$. Additionally, while precipitation appears less important than maximum temperature (\texttt{tmax}) in the period 1990--1994, both current (\texttt{pr\_curr}) and previous year (\texttt{pr\_prev}) precipitation show significant increases in relative importance of these variables between 1995 and 2020, with both overtaking maximum temperature during this period. The covariates windspeed (\texttt{wspd}) and year (\texttt{fire\_yr}) remain relatively unimportant throughout the entire time period. This suggests that dryness is a bigger driving factor of burnt area than high temperatures: this is not surprising, as while high temperatures increase ignition rates (from natural causes) and windspeed affects spread, dryness is {more} directly tied to the total area that has potential for burning. Figure~\ref{fig:ff_VI} also illustrates an overall increase in the VI scores over time, suggesting that these meteorological variables may be becoming increasingly more important. Table~\ref{t:VI} provides the variable importance scores for the shape parameter $\xi$, for each of the three time periods. {The importance scores for the tail largely follow the same trends as the importance scores for the body of the distribution, with short-term and long-term dryness dominating, and \texttt{tmax} showing slight decreases. In particular, \texttt{pr\_prev} shows an eight-fold increase between the 1990--1994 and 2016--2020 periods}. Overall, across the body and the tails, we identify dryness as having the largest effect on the distribution of U.S. wildfire burnt areas.

\begin{table}
\centering
\small
\caption{Estimates of the variable importance scores for the shape parameter, $\xi(\mathbf{x})$, of the SPQRx model fit.}
\begin{tabular}{ccccccc}\toprule
\textbf{Time period} & \texttt{rmin} & \texttt{pr\_curr} & \texttt{pr\_prev} & \texttt{tmax} & \texttt{wspd} & \texttt{fire\_yr} \\\midrule
\textbf{1990--1994} & 2.13 & 0.83 & 0.38 & 1.79 & 0.77 & 0.17 \\
\textbf{1995--2015} & 2.74 & 1.53 & 1.95 & 1.72 & 0.23 & 0.55 \\
\textbf{2016--2020} & 3.29 & 2.35 & 3.01 & 1.17 & 0.45 & 0.04\\\bottomrule
\end{tabular}
\normalsize
\label{t:VI}
\end{table}

\begin{figure}
    \centering
    \includegraphics[width=0.65\linewidth]{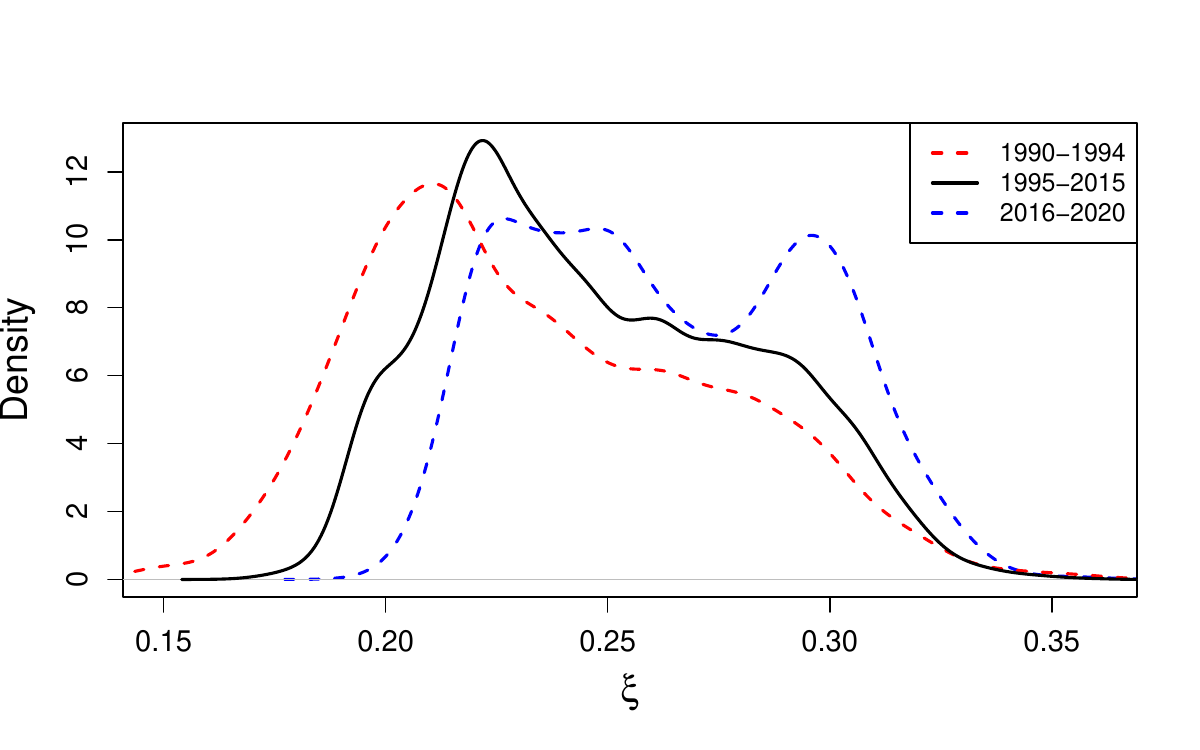}
    \caption{Density of the estimated shape parameter values, $\xi(\mathbf{x})$, stratified by time period.}
    \label{fig:ff_xi_plot}
\end{figure}

Figure~\ref{fig:ff_xi_plot} provides the density of $\xi(\mathbf{x})$ estimates for the training period, 1995--2015, and the two testing periods of 1990--1994 and 2016--2020. Figure~\ref{fig:ff_xi_plot} illustrates a positive temporal trend in the estimates, and suggests that U.S. wildfires are becoming more heavy-tailed, i.e., extreme, over time. We note that the response data have undergone a square-root transformation, and a value of $\xi(\mathbf{x})> 0.25$ on the response scale would translate to approximately $\xi(\mathbf{x})> 0.5$ on the original scale. In the latter case, the blended GP density does not have finite variance, which suggests that the original response data are extremely heavy-tailed.

\begin{figure}[t!]
    \centering
        \includegraphics[width=0.45\linewidth]{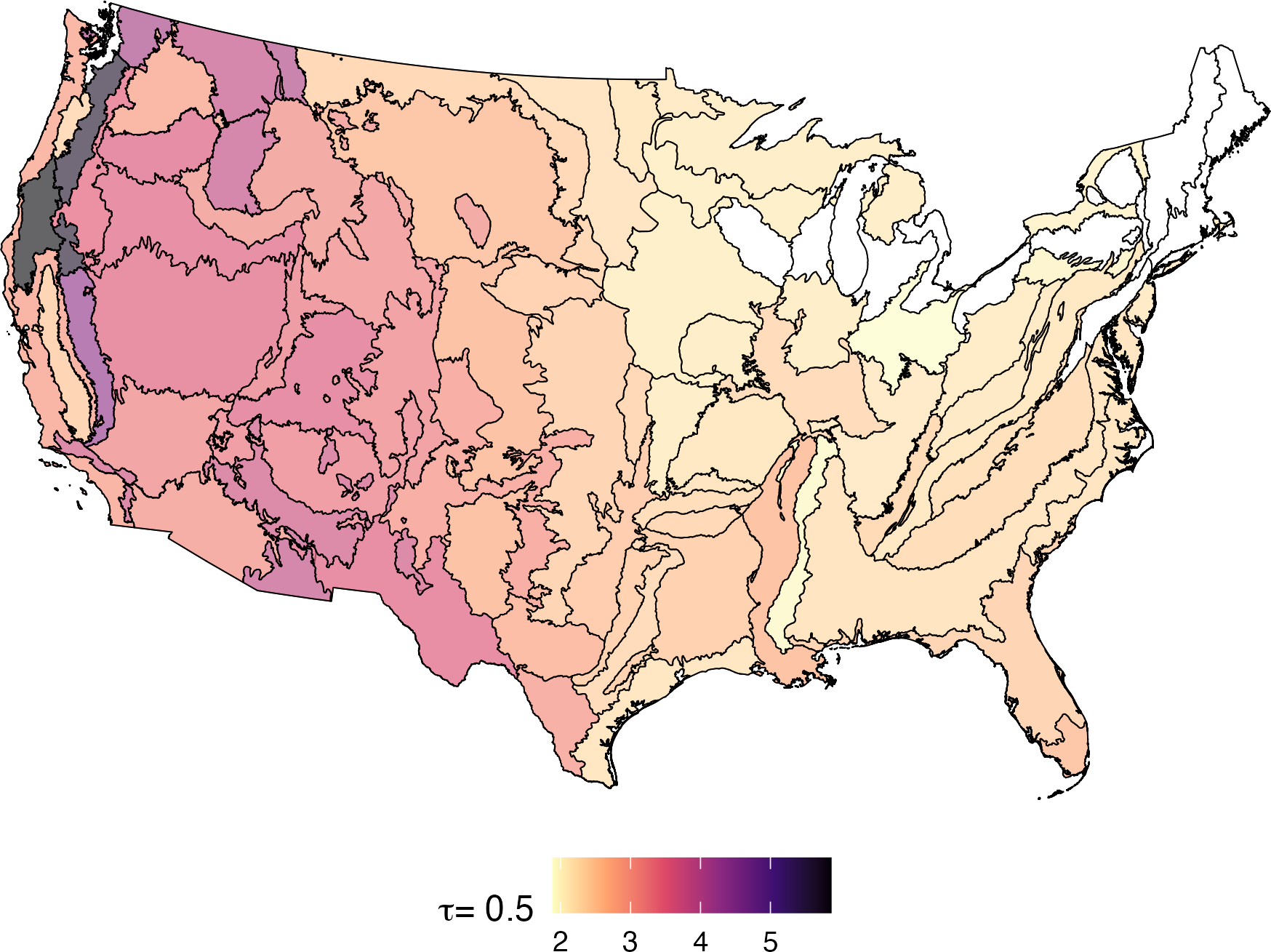}
    \includegraphics[width=0.45\linewidth]{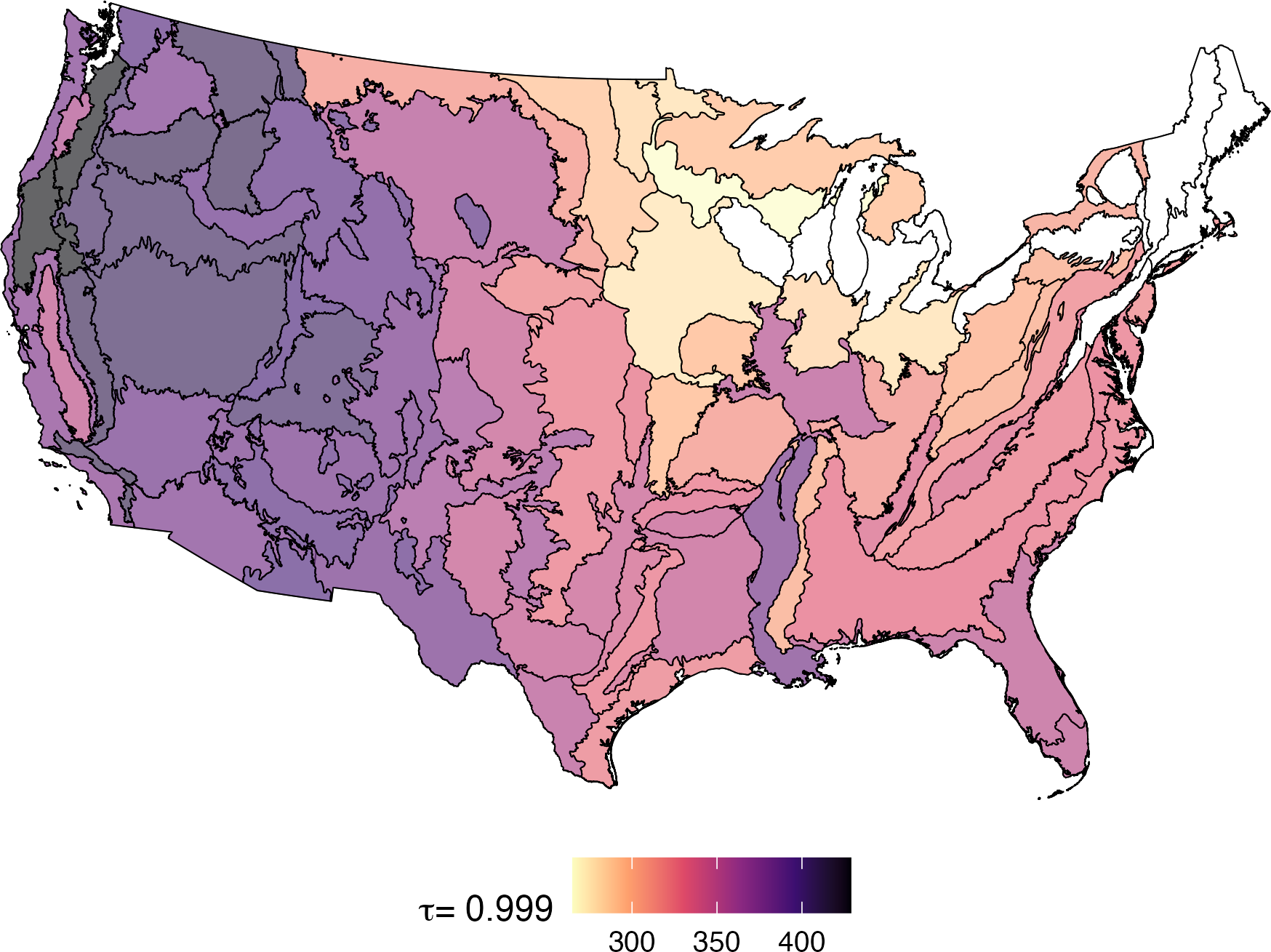}
    \caption{Estimates of the median (left) and 0.999-quantile (right) of burnt area (in 1000s of acres) for all observed wildfires, averaged over each L3 ecoregion. Transparent regions do not include any observed wildfires.}
    \label{fig:xi_l3_map}
\end{figure}
Figure~\ref{fig:xi_l3_map} illustrates the spatial variability in the estimated quantiles of the burnt area response. For each fire in the dataset, we estimate the median and $0.999$-quantile using the fitted SPQRx model. The former quantity is estimated via the SPQR model for the bulk while the latter is estimated via the GP model for the upper-tail. The quantile is squared to obtain the actual burnt area in 1000s of acres, and then aggregated over Level III (L3) ecoregions. The ecoregions group ecosystems that are similar in terms of their biotic, abiotic, terrestrial, and aquatic ecosystem components, including human activity \citep{McMAHON2001,Omernik2014}. Ecoregion boundaries are made available by the U.S. Environmental Protection Agency\footnote{https://www.epa.gov/eco-research/level-iii-and-iv-ecoregions-continental-united-states}, with lower levels indicating more aggregation (and therefore larger ecoregions). 
From Figure~\ref{fig:xi_l3_map}, estimates of the quantiles show spatial variability across the country, with higher values concentrated in the western regions of the U.S. The illustrated east-west gradient is more pronounced in the map of the estimated 0.999-quantile than the median. Alongside the differences in the relative VI scores for the bulk and tail (see Figure~\ref{fig:ff_VI} and Table~\ref{t:VI}), the differences in the spatial patterns of the median and tail quantiles provides further evidence that the covariates have different impacts on the bulk and upper-tail of the wildfire distribution.

\begin{figure}[t]
    \centering
    \includegraphics[width=0.8\linewidth]{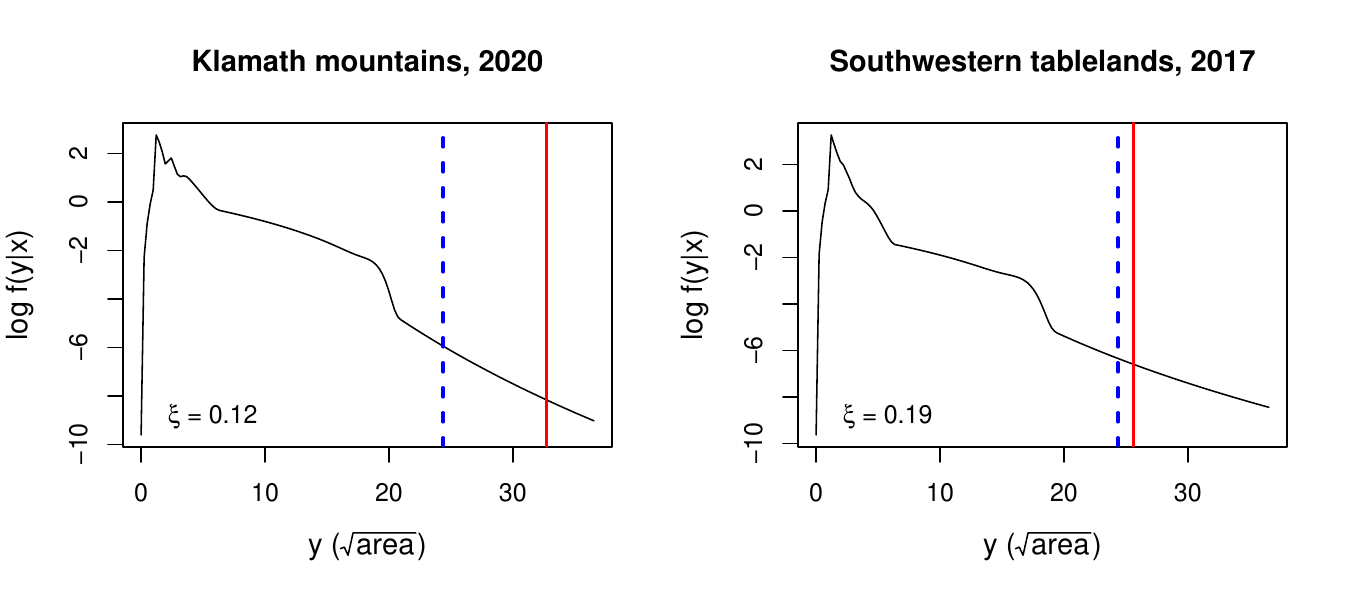}
    \caption{Estimated conditional log-density functions $\log f(y\vert \bx)$ for the two largest wildfires in the dataset. The response variable $Y$ is the square-root of burnt area, originally measured in 1000s of acres. The red lines denote the observed value for the fires, while the dashed blue lines represent the largest value that is observed in the training data.}
    \label{f:bigfire}
\end{figure}

\begin{figure}[t!]
    \centering
    \includegraphics[width=0.8\linewidth]{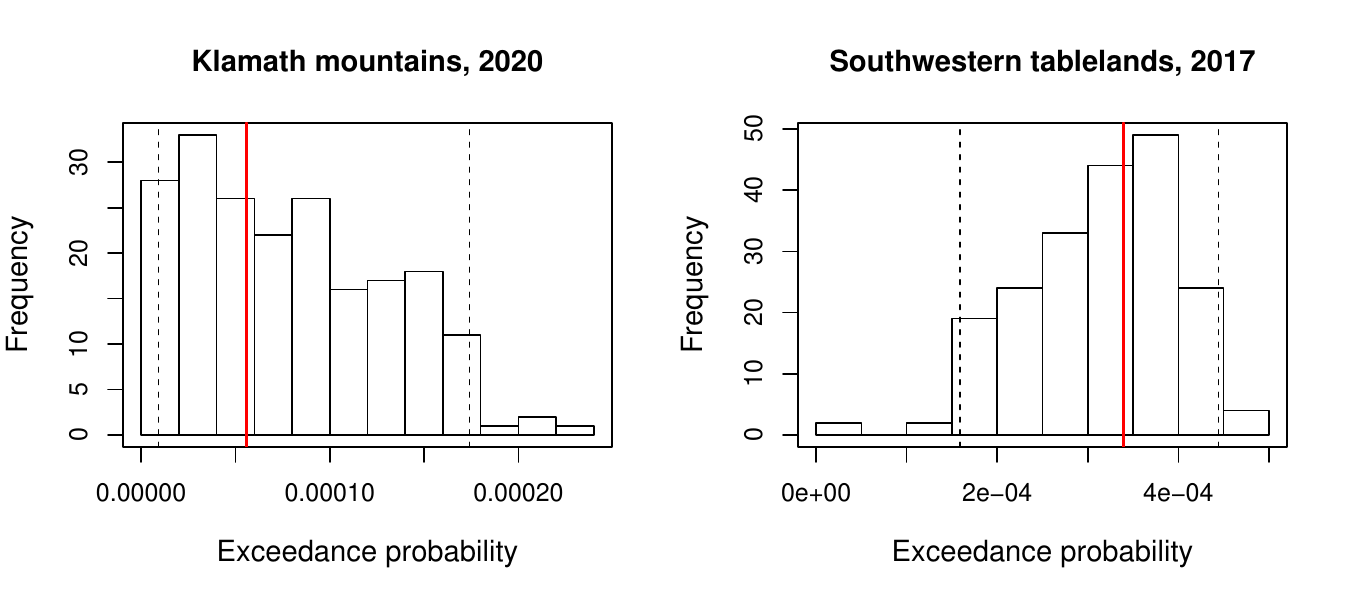}
    \caption{{Bootstrapped estimates of the exceedance probability $\Pr(Y>y\vert \bx)$ for the two largest wildfires in the dataset. The response variable $Y$ is the square-root of burnt area, originally measured in 1000s of acres. The red lines denote the tail probability estimated by SPQRx, while the dashed black lines represent $95\%$ confidence intervals based on 200 bootstrap samples.}}
    \label{f:surv_prob}
\end{figure}

\begin{figure}[b!]
    \centering
    \includegraphics[width=0.8\linewidth]{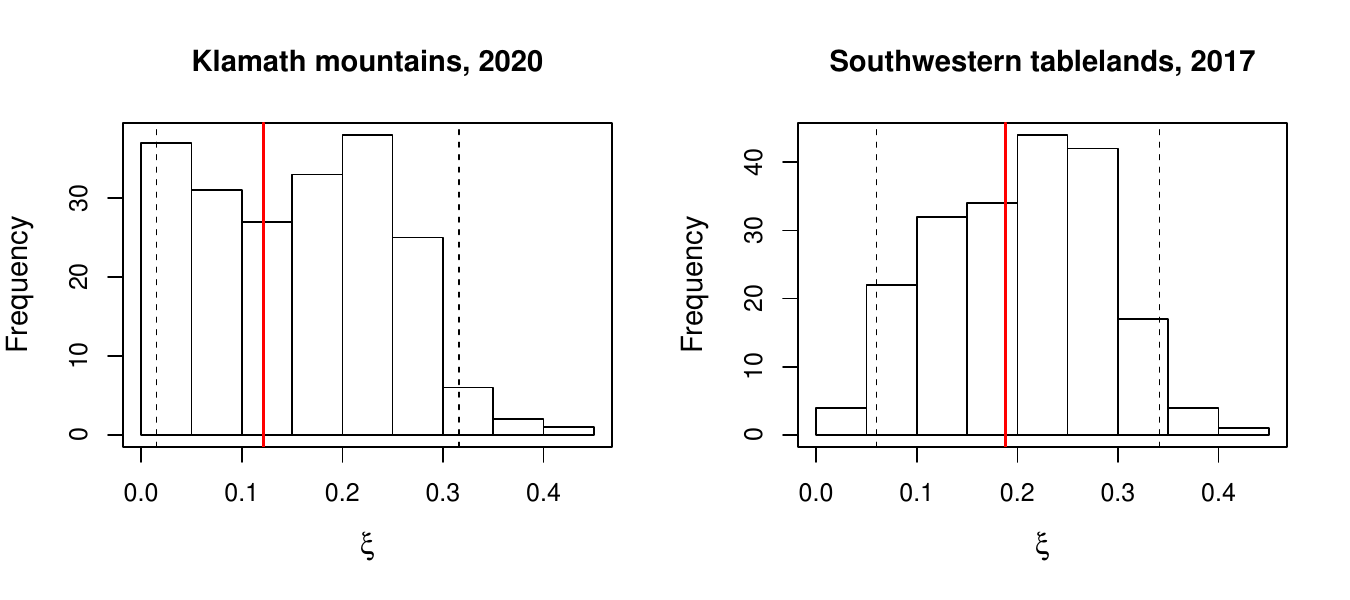}
    \caption{{Bootstrapped estimates of the shape parameter $\xi$ for the two largest wildfires in the dataset. The red lines denote the estimate from SPQRx, while the dashed black lines represent $95\%$ confidence intervals based on 200 bootstrap samples.}}
    \label{f:xi_density}
\end{figure}

Finally, we focus on the two most extreme fires in the dataset (in terms of burnt area). We refer to them as the Klamath mountains and Southwestern tablelands fires, based on their L3 ecoregion of occurrence \citep{lawlershaby2024}. As mentioned above, they are both present in the second test period, 2016--2020, and are therefore outside the range of the training data used to fit our models. Both fires were associated with moderate temperatures ($ <30\degree$C), but with very low {relative} humidity ($<25\%$). Additionally, in the case of the Klamath mountains fire, the aggregate precipitation over the past month (\texttt{pr\_curr}) was in the lowest $0.07$ quantile of the training data. Figure~\ref{f:bigfire} provides the SPQRx estimates of the conditional log-density for these two fires. In each case, the observed wildfires are well beyond the observed maximum of the training data, and are modelled explicitly by the GP upper-tail of the SPQRx model. Figure~\ref{f:bigfire} highlights the flexibility of the SPQRx model which, unlike existing fully-parametric bulk and tail models (see Section~\ref{sec:intro}), is able to capture both the heavy-tails and high kurtosis exhibited by the data. {Figure~\ref{f:surv_prob} plots a histogram of the estimated exceedance probability $\Pr(Y>y\vert \bx)$ for the two fires. The Klamath mountain fire has an estimated exceedance probability of $5.56\times 10^{-5}$ and a $95\%$ confidence interval of $(9.15\times 10^{-6},1.74\times 10^{-4})$. The Southwestern tablelands fire has an exceedance probability of $0.0003$ with a $95\%$ confidence interval of $(0.0001,0.0004)$. Finally, Figure~\ref{f:xi_density} plots uncertainty estimates of $\xi$ for the two fires: the Klamath mountains fire ($\hat{\xi} = 0.12$) has a $95\%$ interval of $(0.02,0.32)$, while the Southwestern tablelands fire ($\hat{\xi} = 0.19$) has a 95\% confidence interval of ($0.06,0.34$). We note that the larger estimate of $\xi(\bx)$ among these two is actually associated with the smaller wildfire. This demonstrates the necessity of a modelling approach like SPQRx that is adept at capturing how covariates affect both the bulk and tail of the burnt area distribution.}
\section{Discussion}
In this work, we developed a flexible framework called semi-parametric quantile regression for extremes (SPQRx) for conditional bulk and tail estimation. To achieve this, we propose the blended generalised Pareto (bGP) distribution, which is a new univariate model for the entire range of data that is EVT-compliant, in that its upper-tail follows exactly a GP distribution. We use the bGP distribution to combine a deep parametric GP regression model with the traditional SPQR framework for conditional density estimation. The flexible nature of SPQR ensures that parametric assumptions do not need to be made about the bulk of the conditional distribution, which can instead be arbitrarily flexible. The SPQRx framework addresses one of the main weaknesses of SPQR by allowing the modelling of heavy-tailed distributions and making predictions outside the range of the observed data. We also obtain variable importance estimates for SPQRx using accumulated local effects. As SPQRx allows covariates to separately affect the bulk and the tail of the distribution, we can perform inference on variable importance for the tails separately of that for the bulk. A simulation study and a real-data application to U.S. wildfire burnt areas demonstrated the efficacy of SPQRx.

{In our current work, the hyper-parameters of the blending function, $(p_a,p_b,c_1,c_2),$ are carefully chosen using numerical experiments and via grid search algorithms. Arbitrarily choosing these hyper-parameters can lead to pathological cases where the density function, $h(y | \mathcal{W}, \xi)$ in \eqref{eq:blendedHdens}, is negative for $y$ in the blending interval, $[a,b]$ (see Section~\ref{sec:gpd}). In practice, we found that this phenomenon was extremely rare, and we mitigated the risk of it occurring during training of the SPQRx model by pre-training (Section~\ref{Ssec:inference}) and penalisation (Appendix~\ref{s:valid}). Future work may investigate the specific constraints required on the constituent distribution functions and weighting function that guarantee validity of the resulting blended GP distribution.}

Another simple extension of the SPQRx framework is to allow for different subsets of variables to affect the bulk and upper-tail of the distribution, which can be achieved by modelling $\mathcal{W}(\mathbf{x})$ and $\xi(\mathbf{x})$ in \eqref{eq:blend_cond_dens} using separate neural networks (with separate inputs). Further work may also consider a blended GP distribution and resulting SPQRx model that allows for extrapolation in both the lower- and upper-tail of the conditional distribution. An extension of \eqref{eq:blendedH}, which blends a constituent model for the bulk with two independent GP distributions for the lower- and upper-tail, is trivial. However, inference for such a model using deep learning may be numerically difficult.

{As noted in Section~\ref{s:spqr_model}, the SPQR representation in \eqref{e:spqrlik} can approximate any continuous density function as the number $K$ of basis functions goes to infinity. Further work could investigate similar theoretical results for SPQRx. As we have equality between the SPQRx and SPQR density and distributions functions on the compact support $[0,a],$ for $a$ the $p_a$-quantile of the SPQR distribution, we might expect a weaker approximation result to hold for SPQRx. A particular case of interest might be that where $p_a\rightarrow 1$, and the SPQRx distribution function converges to its SPQR constituent distribution.}

{Finally, extending SPQRx to a multivariate setting would allow us to study non-stationarity in multivariate distributions, whilst also providing a flexible model which is appropriate for the joint bulk and tails. This would require describing a copula for the bulk using, e.g., tensor splines \citep[][]{ZONG1998341}, alongside an extremal bivariate copula or blended GP marginal tails. Such an approach may provide an alternative to the parametric multivariate bulk and tail models of, e.g., \cite{ANDRE2024107841} and \cite{alotaibi2025joint}. An alternative approach could be to couch SPQRx within the angular-radial modelling frameworks of, e.g., \cite{mackay2024deep} or \cite{murphybarltrop2025}.
}



\if1\blind
{
\subsection*{Funding details} 
The authors have no relevant funding details to disclose.
\subsection*{Acknowledgments}
} \fi
The authors thank Elizabeth S. Lawler for access to the wildfire burnt areas data and for help creating the maps. The authors also thank Brian J. Reich and members of the Glasgow-Edinburgh Extremes Network (GLE$^2$N; \url{glen-scotland.github.io/glen/}) for helpful feedback. This work has made use of the resources provided by the Edinburgh Compute and Data Facility (ECDF) (\url{www.ecdf.ed.ac.uk/}).

\renewcommand{\theequation}{A.\arabic{equation}}
\renewcommand{\thefigure}{A\arabic{figure}}
\renewcommand{\thetable}{A\arabic{table}}
\renewcommand{\thesection}{A\arabic{section}}

\setcounter{figure}{0}
\setcounter{table}{0}
\setcounter{equation}{0}

\spacingset{1.9}

\baselineskip=14pt
\begingroup

\bibliographystyle{apalike}
{\footnotesize 
\bibliography{library}
}
\endgroup
\clearpage

\begin{appendix}
\section*{Appendix}
\section{Construction of $M$- and $I$-splines}
\label{s:spline}
A polynomial spline is a piecewise polynomial function over a compact interval, say $[L,U],$ with the constituent polynomials having the same degree and connecting smoothly at predetermined knots. Several classes of splines, e.g., $B$-splines, also constitute basis functions of their function space \citep{splinesbook}. Similarly, the $M$-spline family of basis functions, $\{M_k\}_{k=1}^K,$ is suited to representing densities \citep{Curry1966}. The $M$-splines are piecewise polynomial functions of order $d$, which are defined on a set of $K+d$ knots, $t_1,\ldots,t_{K+d}$. The basis function $M_k(y)$ is positive on $(t_k,t_{k+d})$, and zero otherwise. Additionally $\int_L^U M_k(y) dy = 1$, which gives $M_k(y)$ a natural interpretation as a density function supported on $[L,U]$. The basis functions are computed using the following system of recursive equations: for $d = 1$,
\[
    M_k(y\vert d) = \begin{cases}
        \frac{1}{t_{k+1} - t_k},\, &t_k \leq y < t_{k+1},\\
        0,\, &\mbox{otherwise}.
    \end{cases}
\]
and, for $d > 1$,
    \begin{align*}
    M_k(y\vert d) = \frac{d[(y - t_k)M_k(y\vert d-1) + (t_{k+d} - y)M_{k+1}(y\vert d-1)]}{(d-1)(t_{k+d} - t_k)}.
\end{align*}
In the case of SPQR, we use an order of $d=3$ for the polynomial basis. We note that $B$-splines are also closely related to $M$-splines, as the $B$-spline basis function is defined by $B_k = (t_{k+d} - t_k)M_k/d$ \citep{ramsay1988}. 

\citet{ramsay1988} also introduced integrated splines, or $I$-splines, over $[L,U]$. These are built using basis functions
\begin{align*}
    I_k(y \vert d) = \int_L^y M_k(u\vert d) \mathrm{d}u.
\end{align*}
As each $M_k(y)$ is a polynomial spline (for $d>1$) that has the properties of a density function, the corresponding  $I$-spline has a natural representation as the corresponding distribution function; $I$-splines are integrals over $M$-splines, and are monotone due to their construction.

Since the $M$-splines form a basis, for a vector of weights, $\mathcal{W} = (w_1,\ldots,w_K)$, the convex combination $\sum_{k=1}^K w_kM_k(y)$ is also an $M$-spline and a valid density function. Therefore, with an appropriate choice of $\mathcal{W}$, the $M$-splines can be used to model univariate density functions in a semi-parametric manner. In the case where covariate information is available, one can allow $\mathcal{W}$ to depend on covariates, and construct a representation for conditional density functions; see Section~\ref{sec:SPQR} of the main text.

\section{Constructing valid SPQRx density functions}
\label{s:valid}
{As discussed in Section~\ref{sec:gpd}, the distribution function defined in \eqref{eq:blendedH} is not guaranteed to be valid for arbitrary combinations of constituent distribution function $F(y|\mathcal{W})$, shape parameter $\xi$, and weight function $p(\cdot)$. When $F(y|\mathcal{W})$ is right-skewed and the constituent GP distribution function has relatively lighter-tails, and $F(y|\mathcal{W})> F_{\rm GP}(y|\mathcal{W})$ for some $y\in [a,b]$, we may find that the blended GP density function is negative for some $y\in [a,b]$. Whilst this is a pathological scenario that is unlikely to occur in practice (and did not affect the aforementioned simulation studies or application), we provide here a short discussion on this problem and practical advice for mitigating the risk of predicting invalid SPQRx distributions during model training. }

\begin{figure}[t!]
\centering
    \includegraphics[width=0.8\linewidth]{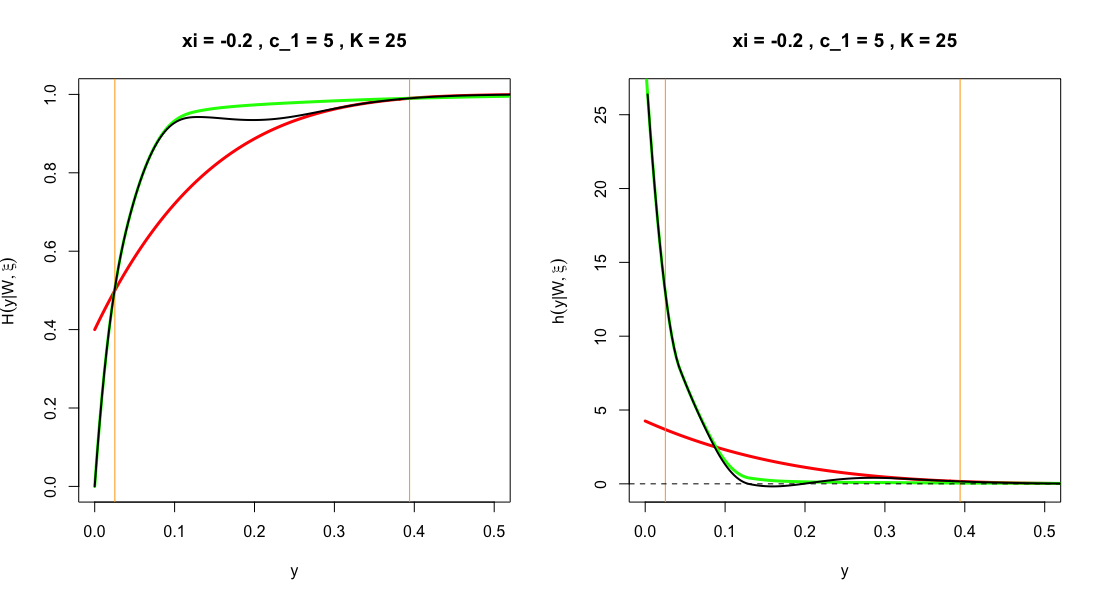}
  \caption{{Illustrative example of an invalid SPQRx distribution (left) and density (right) function. For $K=25$ specified basis coefficients, we construct the SPQR density and distribution functions (green curves). We then find $a=Q_{\rm SPQR}(p_a | \mathcal{W})$ and $b=Q_{\rm SPQR}(p_b|\mathcal{W})$, where here $p_a=0.5$ and $p_b=0.99$. The values of $a$ and $b$ are denoted by the orange horizontal lines, and the dashed horizontal line goes through zero. Then, we find the required GP distribution function (red) to satisfy continuity of the bGP($\mathcal{W},\xi)$ distribution function; its resulting distribution and density functions are provided in black. Here, we have $\xi=-0.2$ and $c_1=c_2=5$. Note that the blended GP density (black, right) is negative for some $y$.}}
    \label{fig:s_example1}
\end{figure}

{Figure~\ref{fig:s_example1} gives an example of an invalid unconditional SPQRx distribution/density function. Here we have chosen $K=25$ monotonically-decreasing basis coefficients that ensure that $f_{\rm SPQR}$ is right-skewed on the interval $[0,1]$. We observe that the bulk of the $f_{\rm SPQR}$ density is at $y$ close to zero, with persistent non-zero mass $f_{\rm SPQR}(y)>0$ as $y$ goes towards its upper-endpoint at one. We take the constituent GP distribution to have shape parameter $\xi=-0.2$, choose the blending interval with levels $p_a=0.5$ and $p_b=0.99$, and set $c_1=c_2=5$ for the weighting function $p(y)$. As Figure~\ref{fig:s_example1} illustrates, we have $F_{\rm SPQR}(y|\mathcal{W})> F_{\rm GP}(y|\mathcal{W})$ for some $y\in [a,b],$ which results in the blended density $h(y | \mathcal{W},\xi)$ being negative for some $y$ in the blending interval. This is a pathological case {that} would not occur in practice: when $\xi$ and $f_{\rm SPQR}$ are jointly estimated, we would expect estimates of $\xi$ to increase to reflect the skewed behaviour of the $f_{\rm SPQR}$ density function. }

\begin{figure}
   \centering
 \includegraphics[width=.78\linewidth]{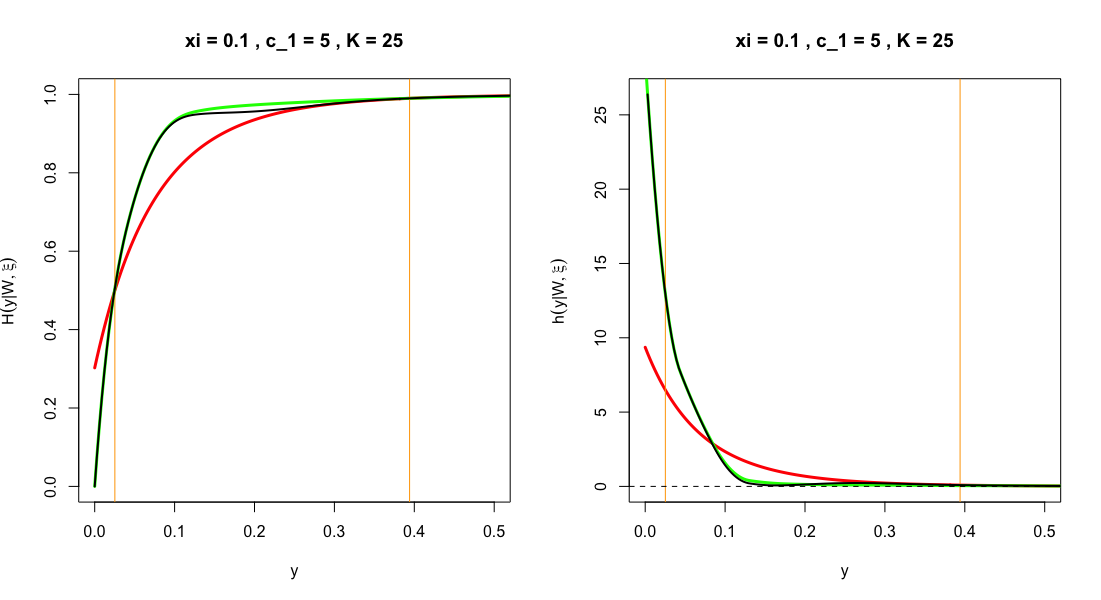}
    \includegraphics[width=.78\linewidth]{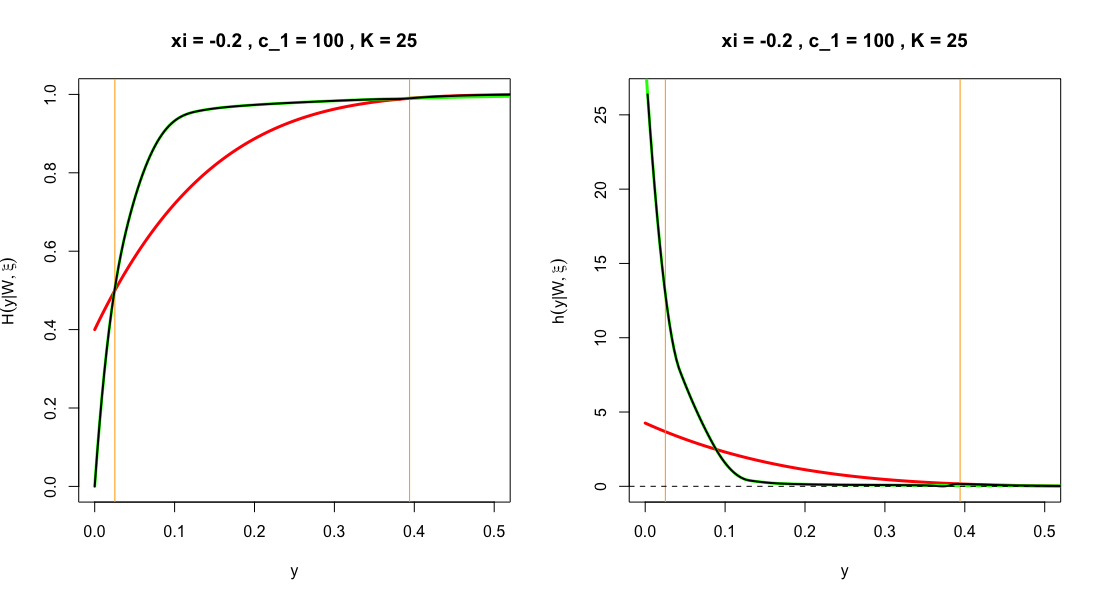}
        \includegraphics[width=.78\linewidth]{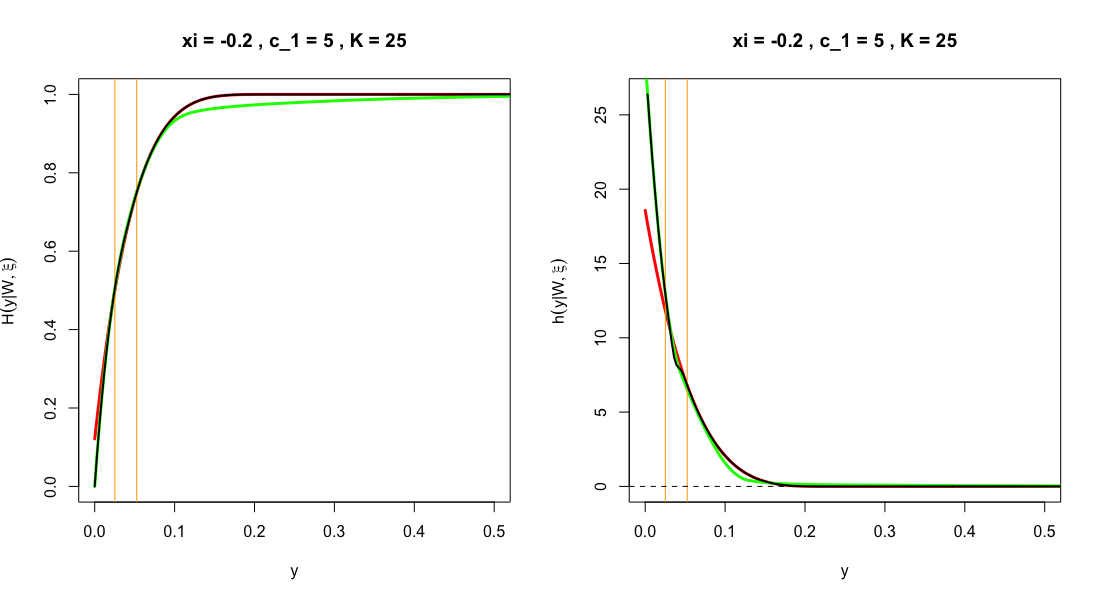}
    \caption{{Illustration of how the SPQRx distribution in Figure~\ref{fig:s_example1} can be adapted to create a valid distribution. Here, we change only one hyper-parameter, relative to Figure~\ref{fig:s_example1}, per row. For the first row, we change $\xi$ from $-0.2$ to $0.1$. For the second row, we change $c_1$ from 5 to 100. For the third row, we change $p_b$ from 0.99 to 0.75. All SPQRx density functions (black) are positive for all $y$. }}
    \label{fig:s_example2}
\end{figure}

{Figure~\ref{fig:s_example2} illustrates three minor changes to the parameterisation of the distribution function in Figure~\ref{fig:s_example1} {that} will create a valid distribution. In the first example, we make the tails of $F_{\rm GP}$ heavier, by increasing $\xi$ to $0.1$. In the second example, we increase $c_2$, relative to $c_1$, which places more weight on the (valid) $F_{\rm SPQR}$ distribution function within the blending interval $[a,b]$. In the third example, we use a smaller blending interval, and lower $p_b$ from $0.99$ to $0.75$. These minor changes all lead to valid distribution functions.} 

{In a practical setting, one can mitigate the risk of predicting an invalid SPQRx distribution function during model training by: i) pre-training the SPQR density function and setting the initial $\xi$ estimate large (see Section~\ref{Ssec:inference}); and ii) penalising invalid density functions. For the penalty, we add to the loss function
$$
\lambda\sum_{\mathbf{x}\in \{\mathbf{x}_i\}_{\rm train}}\int_{0}^1\text{ReLU}\left\{-h(y|\mathcal{W}(\mathbf{x}),\xi(\mathbf{x}))\right\}\mathrm{d}y,
$$
where $\{\mathbf{x}_i\}_{\rm train}$ denotes the training set of covariates, $\lambda > 0$ is the penalty coefficient, and $\text{ReLU}(\cdot) = \max\{0, \cdot\}$ is the rectified linear unit activation function. In practice, we approximate the inner integral using a sequence of $y$ values on $[0,1]$. Throughout the main text and in Appendix~\ref{s:sim_studies}, we use both approaches during training, with $\lambda$ set to 100. In unreported experiments, we find that including the penalty improves the extrapolative performance of SQPRx, particularly when the shape parameter $\xi(\mathbf{x})$ is negative. }
\section{Additional simulation studies}
\label{s:sim_studies}
{We now present two additional simulation studies that complement the one presented in Section~\ref{sec:sim_study} of the main text. We consider a similar setting, with covariates $\mathbf{X}_i, {i=1,\dots,p,}$ drawn independently from a Unif(0,1) distribution. However, we here consider the higher-dimensional setting where $p = 20$ (instead of $p=3$), and where the tail index of the response distribution is allowed to vary with the covariates. In particular, we consider two settings: i) where the response distribution, $Y\mid(\mathbf{X}=\mathbf{x}),$ follows a heavy-tailed unit-Lomax distribution with density function $f(y | \mathbf{x})=\alpha_0(\mathbf{x})(1+y)^{-(\alpha_0(\mathbf{x})+1)}$ for $y>0$ and shape $\alpha_0(\mathbf{x}) >0$; and ii) where $Y\mid(\mathbf{X}=\mathbf{x})$ follows a GP$(1,\xi_0(\mathbf{x}))$ distribution with $\xi_0(\mathbf{x}) \in(-1,0)$ such that the conditional upper-tail is bounded above for all $\mathbf{x}$. For the true parameter functions, $\alpha_0$ and $\xi_0$, we use complex, highly non-linear functions given by
$$
\alpha_0(\mathbf{x})=3+\exp\{-1+\beta(\mathbf{x})\} > 3,\quad\text{and}\quad\xi_0(\mathbf{x})=-1/[1+\exp\{-\beta(\mathbf{x})\}]\in(-1,0),
$$
with 
\begin{align*}
\beta_0(\mathbf{x})&=x_2x_1 + x_6\left[1 - \cos\left(\pi x_3x_4\right)\right] + 2\sin(x_5) / (|x_7 - x_8| + 2)\\
         &+ 0.2 (x_6 + x_8 x_9 / 2) ^ 2 - \sqrt{x_9 ^ 2 + x_{10}^2 + 2}.
\end{align*}
A similar function to $\beta_0$ was previously explored by \cite{richards2022regression} and \cite{Richards2024}. Note that, while $p=20$, only the first ten components of the covariate vector $\mathbf{X}$ act on the response distribution.
}

{
We conduct the two supplementary simulation studies as in Section~\ref{sec:sim_study}. That is, over 100 repeated experiments, we evaluate the out-of-sample IWD and tIWD on a test set of 5000 observations, with models fitted using all considered candidate SPQR, SPQRx, and deep GP models, as described in Section~\ref{sec:sim_study}. The results for settings 1 and 2 are presented in Tables~\ref{stable1} and \ref{stable2}, respectively. We note that, where the data are heavy-tailed (i.e., setting~1), the SPQRx model consistently outperforms SPQR, in terms of lower estimated IWD and tIWD. However, SPQRx provides lower tIWD than the deep GP model in the low sample size setting, where $n = 1000$, suggesting that larger sample sizes may be required to see the benefits of extrapolation using SPQRx instead of standard peaks-over-threshold models. When the conditional upper-tails are bounded above (setting~2, Table~\ref{stable2}), SPQRx is outperformed by SPQR. This is unsurprising as inference for SPQRx can be troublesome when the shape parameter is negative (see discussion in Section~\ref{Ssec:inference}). Both SPQR and SPQRx are outperformed by the deep GP model, in terms of lower tIWD estimates. However, as discussed previously, the deep GP model is unable to provide a description for the full conditional density, and thus cannot provide IWD estimates.
}
\begin{sidewaystable}
\centering
 \caption{{Results of the supplementary simulation study 1, with unit Lomax response distribution. The median of the IWD and tIWD estimates (alongside 50\% confidence intervals; in brackets) are reported for the original SPQR, new SPQRx model (in bold), and the deep GP regression model (underlined). Note that the IWD is undefined for the deep GP model and the hyperparameter $K$ is not used in its construction.  The hyper-parameters of the SPQRx, $(p_a,p_b,c_1)$, are optimised for each row by minimising the tIWD, and are reported in the final column. }}
 \label{stable1}
\begin{tabular}{rrrccr}
  \hline
$n$ & $K$ & $n_h$ & IWD & tIWD & Optimal $(p_a,p_b,c_1)$\\ 
  \hline
\multirow{4}{*}{1000} & 15 & 16 &0.052 (0.035, 0.075)/\textbf{0.042 (0.029, 0.062)} & 3.76 (3.23, 5.27)/\textbf{2.70 (1.46, 5.65)} & $(0.9,0.99,5)$\\ 

& 15 & 32 &0.46 (0.034, 0.072)/\textbf{0.040 (0.031, 0.065)} & 3.75 (3.20, 5.38)/\textbf{3.17 (1.81, 5.94)} & $(0.75,0.99,25)$\\ 

   & 25 & 16 &0.048 (0.038, 0.063)/\textbf{0.040 (0.030, 0.060)} & \shortstack{3.76 (3.21, 5.41)/\textbf{3.00 (2.14, 5.10)} \\ \underline{2.47 (2.14, 3.14)}} & $(0.9,0.99,25)$\\ 
   
   & 25 & 32 &0.045 (0.035, 0.057)/\textbf{0.039 (0.028, 0.058)} & \shortstack{3.76 (3.17, 5.37)/\textbf{3.14 (2.00, 7.49)}\\ \underline{2.46 (2.13, 3.09)}} & $(0.75,0.99,25)$\\ 
   
   \hline
   
  \multirow{4}{*}{10000}  & 15 & 16 &  0.134 (0.078, 0.237)/\textbf{0.132 (0.077, 0.235)} &  3.02 (2.26, 4.95)/\textbf{2.22 (1.40, 3.45)} & $(0.75,0.999,5)$\\ 
  
   & 15 & 32 & 0.134 (0.075, 0.238)/\textbf{0.131 (0.076, 0.237)} & 2.90 (2.21, 4.94)/\textbf{1.95 (1.22, 3.82)}  & $(0.75,0.999,25)$\\ 
   
   & 25 & 16 &0.039 (0.025, 0.066)/\textbf{0.037 (0.025, 0.069)}& \shortstack{4.16 (3.08, 6.94)/\textbf{1.40 ( 0.95, 2.71)}\\\underline{2.18 (2.02, 2.55)}} & $(0.75,0.99,25)$\\
   
   & 25 & 32 &0.039 (0.027, 0.068)/\textbf{0.037 (0.025, 0.070)} & \shortstack{4.43 (2.89, 6.99)/\textbf{1.33 (0.91, 2.51)} \\ \underline{2.20 (1.94, 2.47)}} & $(0.75,0.99,25)$\\ 
   \hline
\end{tabular}
\end{sidewaystable}

\begin{sidewaystable}
\centering
 \caption{{Results of the supplementary simulation study 2, with a bounded GP response distribution. The median of the IWD and tIWD estimates (alongside 50\% confidence intervals; in brackets) are reported for the original SPQR, new SPQRx model (in bold), and the deep GP regression model (underlined). Note that the IWD is undefined for the deep GP model and the hyperparameter $K$ is not used in its construction.  The hyper-parameters of the SPQRx, $(p_a,p_b,c_1)$, are optimised for each row by minimising the tIWD, and are reported in the final column. }}
 \label{stable2}
\begin{tabular}{rrrccr}
  \hline
$n$ & $K$ & $n_h$ & IWD & tIWD & Optimal $(p_a,p_b,c_1)$\\ 
  \hline
\multirow{4}{*}{1000} & 15 & 16 &0.060 (0.056, 0.064)/\textbf{0.059 (0.060, 0.063)} & 0.67 (0.59, 0.87)/\textbf{0.65 (0.59, 0.80)} & $(0.75,0.99,5)$\\ 

& 15 & 32 & 0.059 (0.056, 0.064)/\textbf{0.059 (0.056, 0.064)} &  0.67 (0.59, 0.90)/\textbf{0.67 (0.59, 0.86)} & $(0.75,0.99,5)$\\ 

   & 25 & 16 &0.055 (0.053, 0.058)/\textbf{0.055 (0.052, 0.058)} & \shortstack{0.68 (0.60, 0.89)/\textbf{0.66 (0.60, 0.88)} \\ \underline{0.64 (0.56, 0.73)}} & $(0.75,0.99,25)$\\ 
   
   & 25 & 32 &0.055 (0.052, 0.058)/\textbf{0.054 (0.052, 0.056)} & \shortstack{0.68 (0.60, 0.89)/\textbf{0.66 (0.59, 0.89)}\\ \underline{0.62 (0.55, 0.74)}} & $(0.75,0.99,5)$\\ 
   
   \hline
   
  \multirow{4}{*}{10000}  & 15 & 16 & 0.061 (0.054, 0.064)/\textbf{0.061 (0.054, 0.063)} &   0.66 (0.62, 0.75)/\textbf{0.56 (0.50, 0.63)} & $(0.9,0.99,25)$\\ 
  
   & 15 & 32 & 0.061 (0.054, 0.067)/\textbf{0.060 (0.053, 0.063)} &0.70 (0.65, 0.78)/\textbf{0.59 (0.53, 0.65)}  & $(0.75,0.99,25)$\\ 
   
   & 25 & 16 &0.047 (0.042, 0.051)/\textbf{0.044 (0.042, 0.048)}& \shortstack{0.73 (0.65, 0.82)/\textbf{0.61 (0.57, 0.69)}\\\underline{0.41 (0.36, 0.44)}} & $(0.75,0.99,25)$\\
   
   & 25 & 32 &0.046 (0.046, 0.051)/\textbf{0.046 (0.043, 0.047)} & \shortstack{0.74 (0.68, 0.81)/\textbf{0.62 (0.57, 0.67)} \\ \underline{0.33 (0.31, 0.35)}} & $(0.75,0.99,25)$\\ 
   \hline
\end{tabular}
\end{sidewaystable}
\end{appendix}

\end{document}

%% file: commands.tex
\newcommand{\bb}{ \mbox{\bf b}}

\newcommand{\bx}{ \mbox{\bf x}}
\newcommand{\bX}{ \mbox{\bf X}}

\newcommand{\bW}{ \mbox{\bf W}}

\newcommand{\beq}{ \begin{equation}}
\newcommand{\eeq}{ \end{equation}}
\newcommand{\beqn}{ \begin{eqnarray}}
\newcommand{\eeqn}{ \end{eqnarray}}

\usepackage{caption}